\documentclass[12pt]{article}
\usepackage{amsfonts}
\usepackage{amsmath}
\usepackage[labelfont=bf]{caption}
\usepackage{float}
\usepackage{footmisc}
\usepackage{graphicx,datetime}
\usepackage{geometry}
\usepackage{hyperref}
\usepackage{mathtools}
\usepackage{natbib}
\usepackage{tikz}
\usepackage{setspace}
\usepackage{xcolor}

\usetikzlibrary{calc}

\newcommand*{\Resize}
[2]{\resizebox{\textwidth}{!}{$#2$}}

\newtheorem{theorem}{Theorem}

\newtheorem{claim}{Claim}

\newtheorem{lemma}{Lemma}

\newtheorem{proposition}{Proposition}

\newenvironment{proof}[1][Proof]{\noindent\textbf{#1.} }{\ \rule{0.8em}{0.8em}}
\hypersetup{
colorlinks,
linkcolor={red!50!black},
citecolor={blue!50!black},
urlcolor={blue!80!black}
}
\begin{document}

\title{Information Acquisition and Diffusion in Markets\thanks{{\small {\ We
thank seminar participants at the Vienna Graduate School of Economics, EARIE
2017, EEA-ESEM 2018, and the 2018 Consumer Search and Switching Costs
Workshop especially Daniel Garcia, Eeva Mauring, Mariya Teteryatnikova,
Sandro Shelegia, John Vickers and Chris Wilson for helpful suggestions and
comments on earlier versions of the paper. Atayev acknowledges
financial support from the uni:docs Fellowship Programme. Janssen
acknowledges financial support from the Austrian Science Foundation FWF
under project number I 3487.}}}}
\author{Atabek Atayev\thanks{{\small {\ ZEW---Leibniz Centre for European 
Economic Research in Mannheim. Email: atabek.atayev@zew.de}}} \quad
and \ \ \ Maarten Janssen\thanks{{\small {\ Department of Economics,
University of Vienna, National Research University Higher School of
Economics and CEPR. Email: maarten.janssen@univie.ac.at}}} \quad }
\date{\today\\
}

\begin{singlespace}
	\maketitle

\begin{abstract}
	\noindent Consumers can acquire information through their own search 
	efforts
	or through their social network. Information diffusion via word-of-mouth
	communication leads to some consumers free-riding on their ``friends" and
	less information acquisition via active search. Free-riding also has an
	important positive effect, however, in that consumers that do not 
	actively
	search themselves are more likely to be able to compare prices before
	purchase, imposing competitive pressure on firms. We show how market 
	prices
	depend on the characteristics of the network and on search cost. For
	example, if the search cost becomes small, price dispersion disappears,
	while the price level converges to the monopoly level, implying that
	expected prices are decreasing for small enough search cost. More 
	connected
	societies have lower market prices, while price dispersion remains even 
	in
	fully connected societies.
	
	\bigskip
	
	\noindent \textbf{JEL Classification}: D43, D83, D85
	
	\noindent \textbf{Keywords}: Consumer Search; Word-of-Mouth 
	Communication;
	Social Networks.
	
\end{abstract}
	
\end{singlespace}

\sloppy

\newpage

\section{Introduction}

\label{sec:intro}

Decentralized markets rely on how information that is dispersed over many
individuals is diffused (\cite{hayek1945}). Individual agents are, however,
not endowed with a natural amount of information. Often, they have to spend
resources, such as time, to search and acquire information. Accordingly,
agents will only acquire information if the expected benefits exceed the
opportunity cost of doing so. This has led \cite{grossmanstiglitz1980} to
pose that efficient markets cannot exist if arbitrage is costly. Information
can, however, also be acquired in less costly ways, namely through
word-of-mouth (WOM) communication via friends (see, e.g., \cite%
{ellisonfudenberg1995} and \cite{campbell2013}).\footnote{\cite%
{katzlazarsfeld1955} is the classic study showing that information acquired
through personal contacts is the prime reason why people buy a product.} WOM
communication may come with a delay, however, as one has to wait for friends
to communicate their information.

Costly information acquisition and diffusion (WOM communication) are clearly
related. When few people acquire information themselves, little information
will be diffused, while if information is disseminated efficiently, people
may not have the incentive to spend resources to acquire information
themselves. Thus, it is important to understand the interaction between the
incentives to acquire information and the efficiency of the information
diffusion process. This is especially so for online markets and online
interaction through social networks, such as Facebook or LinkedIn. It is
well documented\footnote{%
See, e.g., \cite{browngoolsbee2002}, \cite{jensen2007} \cite{aker2010}, and 
\cite{akermbiti2010}. On word-of-mouth communication, see, e.g., \cite%
{godesmayzlin2004}, \cite{chenetal2011} and \cite{seileretal2019}.} that
online technologies have significantly reduced the search cost related to
information acquisition and increased the possibilities of diffusion and it
is important to understand how these developments affect market outcomes.

In this paper we study the interaction between information acquisition,
diffusion and market power, and explain the impact of changes in the
connectedness of people (impacting diffusion of information through WOM) and
search costs on market outcomes. We adopt a simple theoretical framework of
a homogeneous goods market where firms set prices and consumers engage in
costly sequential search to acquire information about prices before buying
one unit of the good (\cite{diamond1971} and \cite{stigler1961}).\footnote{%
Undoubtedly, in real world markets consumers also exchange information about
product characteristics, such as quality, appearance and/or convenience.
Yet, people also share information online about prices. In homogeneous goods
markets, price communication is the only thing that matters.} Consumers that
have searched for prices spread this information through their network. The
environment we study allows us to consider the impact of social networks and
search costs on information acquisition and market power. In our model, a
network is characterized by two features: (i) the number of friends
different consumers are connected to, and (ii) the speed with which
information flows through the network.

We find that independent of the acquisition (search) cost there always
exists a no-trade equilibrium that is similar to the Diamond paradox: as no
one will acquire information if firms set very high prices and setting high
prices is optimal if no one acquires information.\footnote{%
A no trade equilibrium exists in many simultaneous and sequential search
models where the first search is costly for all consumers (see, e.g., \cite%
{burdettjudd1983} and \cite{diamond1971})} Importantly, WOM resolves the
Diamond paradox in that it creates additional equilibria with positive
sales. in all of these equilibria, searching consumers follow a reservation
price strategy and firms do not price above this reservation price. Under
WOM, consumers determine whether or not to acquire information themselves.
The possibility to get information through their social network implies that
in any equilibrium it cannot be the case that all consumers acquire
information themselves. If that were the case, prices would be equal to the
willingness to pay and consumers would not be willing to spend a positive
search cost so that free riding on also allows them to buy, possibly at the
lowest price charged in the market without incurring the search cost. Thus,
an endogenously determined fraction of consumers is informed through
friends. As by \textit{free-riding on the information acquisition of friends 
} the non-searching consumers will be informed with positive probability
about different prices, it has an important \textit{positive effect }on
market competition.

In terms of the impact of the social network structure, we show that,
contrary to what one may expect, even when the network gets very dense and
many consumers have many connections, prices do \textit{not }converge to
marginal cost and price dispersion remains. What matters for price
dispersion and also for the expected market price is the relative fraction
of consumers that is informed of only one price. This fraction is
endogenously determined and remains positive even in dense networks as
information from friends comes with a delay creating an incentive to search
and searching consumers buy immediately after acquiring information.

The speed of information diffusion in networks is important in that it is a
key determinant of the cost associated with waiting for information through
WOM. We show that a higher speed of information diffusion has two opposing
effects on market competition. First, it has a direct positive effect as
more consumers do not search themselves, making price comparisons more
likely. There is also an indirect effect, however, namely that as prices and
price dispersion decline, consumers have more incentives to become active
searchers themselves, especially when the speed of information diffusion is
low to begin with. These two effects lead both the fraction of active
consumers and firms' profits to have an inverted U-shape with respect to the
speed of information diffusion: when this speed is low to begin with, firms
have an incentive to increase it as this will also speed up their sales, but
when the speed is already relatively large, the effect on expected price
dominates and firms want to slow down information diffusion.

The impact of search cost is best illustrated by considering the case where
the search cost becomes arbitrarily small. We show that in this case price
dispersion disappears and almost all consumers become active themselves and
buy immediately after they have searched themselves. As almost no consumer
makes price comparisons, prices converge to monopoly levels. When the search
cost increases, more consumers remain passive so that a larger fraction of
consumers make price comparisons, resulting in lower prices. Thus, and
contrary to common wisdom, prices are decreasing in search cost when WOM
plays a role.

Our paper provides a new argument to overcome the Diamond paradox.\footnote{%
If individual consumers have downward sloping demand (or the first search is
somehow free), then the Diamond paradox takes on a somewhat different form,
namely that all firms charge the monopoly price.} \cite{wolinsky1986}
resolves the Diamond paradox by having firms produce heterogeneous products
and consumers searching not only for price bu also for a good product match. 
\cite{stahl1989} imposes search cost heterogeneity among consumers, where
some exogenously determined fraction of \textquotedblleft shoppers" have
zero search cost and compare all prices before buying. Unlike these two
papers, we endow consumers with the possibility to acquire information
through WOM in addition to their own information acquisition by means of
search. In this way, we endogenize the fraction of price comparing consumers
and resolve the Diamond paradox.

\cite{galeotti2010} also combines WOM communication and consumer search.
There are three main modeling differences between his paper and ours.\
First, in \cite{galeotti2010} consumers search for prices in a
non-sequential fashion, whereas we have a sequential search framework. In
most consumer retail markets, consumers observe the price at a firm before
they decide whether to search another firm, making the sequential search
paradigm more relevant. Second, \cite{galeotti2010} assumes the first search
is free so that all consumers know at least one price, whereas we have truly
costly search. Finally, where all consumers have the same number of links in
the basic model studied in \cite{galeotti2010}, we model the social network
as a random graph. These differences in modeling lead to pronounced
differences in our understanding of information acquisition and diffusion in
markets. First, in our setting prices are increasing in search cost and tend
to the monopoly price when search cost tends to zero, whereas in\ Galeotti
prices converge to their competitive levels. Sequential search is the main
reason for this difference. Second, if people get better connected and the
network becomes dense, we show that price dispersion and market power
remain, whereas in \cite{galeotti2010} prices converge to marginal cost.
Costly first search is responsible for this important difference.\footnote{%
\cite{janssenetal2005} were the first to study the impact of the first
search being costly on the participation of consumers in the marketplace. In
their setting (and in contrast to ours), search is, however, the only source
of information acquisition.} Third, our result that free-riding has a
positive impact on prices is not present in \cite{galeotti2010} as under
non-sequential search with the first search being free, searching consumers
are always better informed than non-searching consumers.\footnote{\cite%
{miegielsen2014} adopts a sequential search framework but considers a model
where somehow consumers possess information about prices before engaging in
search and the amount of information that consumers have (and share with
each other) is given exogenously.} Fourth, our modeling of search and
communication through a network allows us to consider the impact of the
speed with which communication travels through the network. Finally, we
offer a new resolution to the Diamond paradox.

There is also a growing literature on how WOM communication affects the
pricing and advertising policy of firms in the market. Earlier papers in
this literature (see, e.g., \cite{arbatskayakonishi2016}, \cite%
{biyalogorskyetal2001}, \cite{bloch2016}, \cite{campbell2013}, \cite%
{chuhay2015}, \cite{fainmessergaleotti2016}, \cite{galeottigoyal2009}, \cite%
{junkim2008}, and \cite{kornishqiuping2010}) consider how a monopoly firm
can introduce its product optimally through a network assuming that
consumers passively wait until they receive an advertisement from the firm,
or they are informed through their network. Instead, we allow consumers to
actively reach out and search for information and we study markets where
firms compete in prices. A more recent literature on networks concerns
strategic interaction of competitors on networks. \cite{chenetal2018} and 
\cite{fainmessergaleotti2020} study how firms price discriminate between
different individuals depending on whether or not an individual influences
many consumers. \cite{campbell2019} studies how in markets with product
differentiation awareness of the available products is communicated through
a social network and how this affects firms' prices and the efficiency of
market outcomes. \cite{campbelletal2020} consider how information about
product quality of an experience good flows through the network and how this
affects the quality provision by firms. None of these papers studies,
however, the interaction between the incentives of consumers to acquire
their own information through search and the diffusion of information
through the social network and their impact on firms' prices and market
outcomes.

The rest of the paper is organized as follows. The next section presents the
baseline model, while Section 3 examines markets where consumers and firms
are symmetrically (un)informed about the network structure. Section 4
provides the comparative statics analysis. Section 5 shows how the
equilibrium outcome of the baseline model can be generated in a more general
context. Section 6 continues the analysis by examining markets where
consumers and firms are asymmetrically informed about the network: consumers
know from how many friends they may acquire information before engaging in
search activities, while firms only know aggregate network characteristics.
We conclude with a discussion.

\section{The Baseline Model and Preliminary Results}

\label{sec:Model}

We consider a duopoly\footnote{%
With more than two firms, the characterization of the mixed strategy
distribution in prices is more complicated and in this case, it is difficult
to analyze the gains of search versus the gains of free riding. \cite%
{galeotti2010} also considers duopoly markets.} market for a homogeneous
good where firms compete in prices. The unit cost of production is constant
and normalized to zero. As firms may choose mixed strategies, we denote the
strategy of a firm $i$ by $F_{i}(p)$, representing the probability that a
firm charges a price not larger than $p$. The support of the price
distribution is determined endogenously with $\underline{p}$ and $\overline{p%
}$ being the lower and upper bound of the support, with the possibility of
some prices in the interior of the interval not being chosen.

On the demand side of the market, there is a countably infinite number of
consumers, normalized to one, each with unit demand and a willingness to pay
equal to $v>0$. A consumer buying at price $p$ receives a pay-off of $v-p.$
If a consumer does not consume, she receives a payoff of zero. The
consumers' choice situation is based on the sequential search paradigm (see,
e.g., \cite{kohnshavell1974} and \cite{stahl1989}), where after obtaining
one price quote at a search cost $s,$ with $0<s<v$, consumers have to decide
whether or not to continue to search. We add that because of WOM
communication, consumers can also get information through their network of
friends. This implies that we have to specify in detail how these two
sources of information are intertwined. There are different ways to do this,
but they all share the basic idea that at any point in the search process,
consumers have the choice between actively searching themselves for price
information from firms (which, like in the consumer search literature,
involves a search cost), buying (which is only possible after being informed
about at least one price), and waiting for friends to provide them with
information (or not searching). The latter, we see as a period of time that
the consumer is not active in the (online or offline) product market. In
that period of time, one or multiple friends may have sent messages, and
pay-offs come wih a time delay.

In this section we present a reduced form model with two periods to analyze
this interaction. At the outset, in period 1, consumers individually and
simultaneously decide whether to search or not. A consumer who decides not
to search herself in period 1 cannot buy and has to wait until period 2.
Searching consumers observe a price and can decide to immediately buy, or
continue to search in the same period. The first search is random and costly
and search is with perfect recall. Consumers who acquired information share
this with their immediate neighbors in the social network. In the beginning
of period 2 the non-searching consumers become active online, observe the
information their friends have shared and decide whether to buy based on
that information or to search themselves. Second-period pay-offs are
discounted by a factor $0<\delta <1$.

The social network through which information is diffused is modeled by a
given random graph. The probability a consumer has $k$ links is denoted by $%
t(k)$, $k\in \left\{ 1,2,...\right\} =O$ with $\sum_{k\in O}t(k)=1$. We
denote by $q(k)$ the fraction of consumers with $k$ friends who choose to
search themselves. In the next two sections, we first consider the case,
however, where consumers do not or cannot condition their decision of
becoming active searchers on the number of connections, and then $q$ is
independent of $k.$

The timing of decisions is as follows. First, firms simultaneously set
prices. Second, the network structure is realized (so that firms cannot
condition their pricing strategy on the details of the network structure).
Third, not knowing the prices, consumers simultaneously choose their
strategies as described above.

A few comments on the interpretation of the model are in order. First, the
reduced form interaction between search and word-of-mouth is arbitrary in
many ways, e.g., in that it only allows for two periods, that a searching
consumer cannot decide to not continue to search and wait for information
from friends and that consumers only receive information from direct friends
in the network so that information decays after one step. In\ Section 5 we
show that provided that the search cost is small enough, the equilibrium
outcome of this reduced form model is identical to the equilibrium outcome
of a more general multi-period model where in every period consumers may
decide to buy using the information they have, search themselves, or not
search (and go to the next period where they may receive more information
from friends). We also discuss that our qualitative results are not affected
if information flows to everyone who is connected. Second, we take the point
of view that consumers engage in their social network for many reasons, not
only to exchange price information through friends. Thus, the social network
is given and we do not study the incentives to form links. Third, we think
of $\delta \ $as a measure of the speed of information communication. If $%
\delta $ is high, communication is fast and non-searching consumers quickly
obtain information from searching friends. What is important is that there
is some time elapse before a non-searching consumer becomes active (online)
again and that within that period she may have obtained information from
multiple friends. The advantage of receiving information via friends instead
of actively searching oneself is to economize on search cost. A consumer can
simply follow a referral and purchase without incurring a cost.\footnote{%
Alternatively, one could decompose the full cost of searching for and buying
from a firm into two parts: a true cost of search and a cost of buying the
good. In that case, following a referral consumers still have to incur the
buying cost. In \cite{atayevjanssen2019} we show that the analysis will be
largely unaffected as the buying cost effectively reduces the willingness to
pay.} Fourth, following the literature on observational learning (see, \cite%
{kircherpostlewaite2008}, \cite{garciashelegia2018}), we also could have
assumed that consumers can credibly exchange information only after having
purchased goods. In \cite{atayevjanssen2019} we show that our analysis
applies equally to this case.

We use symmetric perfect Bayesian equilibria (PBE) as solution concept. A
PBE is described by a set of firms' and consumers' strategies such that each
is choosing optimally given beliefs and the strategies of the others.

In the remaining of this section, we show two preliminary results. First,
any PBE where trade occurs is a so-called reservation price equilibrium,
where consumers buy if, and only if, the price they observe is below a
threshold price. Denote by $r$ the reservation price at which consumers are
indifferent between buying immediately and continuing to search. As
consumers' pay-offs of buying and continuing to search after observing a
price $\widetilde{p}$ are given by $v-\widetilde{p}$ and $v-(1-F(\widetilde{p%
}))\widetilde{p}-F(\widetilde{p})E(p|p<\widetilde{p})-s$, respectively, it
is easy to see that $r$ is implicitly defined by 
\begin{equation}
F(r)(r-E[p|p\leq r])=s.  \label{eq:stoppingrule}
\end{equation}%
If there is a solution to (\ref{eq:stoppingrule}), it must be unique for any
non-degenerate $F(p).$

The following lemma rules out all equilibria where after receiving some
information, consumers decide to continue to search. Thus, a searching
consumer will always immediately buy if the price is smaller than $v$ and
non-searching consumers who have received price information from friends
will not decide to search themselves.

\begin{lemma}
\label{lem:F(r)} Any symmetric equilibrium where goods are bought has $%
F(r)=1 $ .
\end{lemma}

This lemma is important in that it rules out equilibria where firms charge
prices such that it is in the consumers' interest to continue to search. The
result is familiar from the literature on sequential consumer search: if an
equilibrium with prices $p$ larger than $r$ would exist, a firm's price will
always be compared with another price, which incentivizes a firm to undercut
the price of the competitor and not to set the largest price in the support
of $F(p)$.

The second preliminary result is that there always exists a trivial ``no
trade" equilibrium.

\begin{lemma}
For any $s>0$, there exists an equilibrium without sales where $q(k)=0$ for
all $k$.
\end{lemma}

As the first search is costly, this not surprising: knowing they will not
sell to anyone, firms may set prices larger than $v-s$ and this pricing
behavior rationalizes consumers' beliefs that it is not rational to search.
If no one searches, no information is shared and consumers cannot buy.

\section{Active Markets}

\label{sec:active markets}

In this section and the next, we focus on symmetric equilibria with positive
sales, i.e., $0<q<1.$ As $F(r)=1,$ all searching consumers make a purchase
at the first search so that a searching consumer's expected payoff is equal
to 
\begin{equation*}
v-E[p]-s.
\end{equation*}

Given that non-searching consumers will always follow the information they
are provided with and will search in period 2 if they did not receive
information from friends, their expected payoff is given by 
\begin{equation*}
\Resize{}{ \delta\sum\limits_{k \in
O}t(k)\left[v-\sum\limits_{m=1}^{k}\binom{k}{m}q^m(1-q)^{k-m}\left(%
\frac{1}{2^{m-1}}E[p] - \left(1-\frac{1}{2^{m-1}}\right) E_{\min
}[p]\right)- (1-q)^k(E[p]+s)\right]. }
\end{equation*}%
This expression can be understood as follows. A non-searching consumer
always makes a purchase either because she gets informed about a price from
friends or she searches herself. The probability that $m$ of a consumer's $k$
friends search is equal to $\binom{k}{m}q^{m}(1-q)^{k-m}$. They all visit
the same firm with probability $1/2^{m-1},$ in which case she buys at the
expected price. When the friends happen to search different firms, which
happens with probability $1-1/2^{m-1}$, the consumer pays the lowest of the
two prices, which in expected terms is $E_{\min }[p]$. The probability that
none of a consumer's $k$ friends search is $(1-q)^{k}$, in which case the
consumer searches herself incurring $s$ and buys from the first visited firm
paying $E[p]$. The entire payoff depends on how fast information arrives to
the consumer, and thus on the discount factor $\delta $.

To be able to handle these expressions, it is useful to introduce for any
arbitrary $0\leq x\leq 1,$ the function $\tau (x)=\sum_{k\in O}t(k)x^{k},$
as the probability generating function. Two expressions will be of
particular importance in our analysis: with if each consumer searching with
probability $q$, $\tau (1-q)$ represents the probability that a consumer has
only friends that do not search themselves, whereas $1-\tau \left( 1-\frac{q%
}{2}\right) $ represents the conditional probability that a consumer who has
obtained one price quote already obtains information concerning the
competitor's price through the network of friends. It follows that

\begin{equation*}
\sum_{k=1}^{N}t(k)\sum_{j=0}^{k}\binom{k}{j}q^{j}(1-q)^{k-j}y^{j}=\tau
(qy+(1-q)).
\end{equation*}%
As $\tau (x)$ is a convex function with $\tau (1)=1$ and $1-\frac{q}{2}%
=(1-q+1)/2$, it follows that $\widetilde{\tau }(q)\equiv 1+\tau (1-q)-2\tau
\left( 1-\frac{q}{2}\right) >0$ for all $q>0$; $\widetilde{\tau }(q)$ is the
ex ante probability that a consumer's friends have observed two different
prices. Also, $\tau ^{\prime }(1)=\lim_{q\rightarrow 1}\tau
(1-q)/(1-q)=t(1). $

The probability $q$ is determined such that the consumer is indifferent
between searching and not searching and, therefore, in equilibrium the above
two expressions have to be equal. Using the probability generating function,
the indifference condition can be written as 
\begin{equation}
(1-\delta )(v-E[p])=(1-\delta \tau (1-q))s+\delta \widetilde{\tau }(q)\left(
E[p]-E_{\min }[p]\right) .  \label{eq:IC_after}
\end{equation}

We now turn to the determination of the equilibrium pricing strategy of the
firms. Setting price $p,$ an individual firm's expected profit is given by 
\begin{equation*}
\Resize{}{ \Pi(p) = \left(\dfrac{q}{2} + \delta(1-q)\left\{\sum\limits_{k
\in O}t(k) \sum\limits_{m=1}^k\binom{k}{m}q^m(1-q)^{k-m}
\left[\dfrac{1}{2^m} + \left(1 - \dfrac{1}{2^{m-1}}\right)
(1-F(p))\right]\right\}+\dfrac{(1-q)^k}{2}\right)p. }
\end{equation*}%
Clearly, a consumer is active with probability $q$ and half of the times she
visits the firm under question to buy outright. With probability $1-q$, a
consumer is passive in which case she purchases only after the information
arrives to her, thus speed of information diffusion $\delta $. She
definitely buys from the firm if all of her $m$ (out of $k$) active friends
happen to visit the firm, which happens with probability $1/2^{m}$. With
probability $1-1/2^{m-1}$, these active friends happen to visit different
firms, in which case the firm under question makes sales only if its price
is lower than that of the rival firm, or $1-F(p)$. Finally, with probability 
$(1-q)^{k}$ none of her friends search, in which case a consumer searches
herself and buys from the firm half of the time. Using the probability
generating functions, the profit expression can be rewritten as follows: $%
E_{\min }[p]$ 
\begin{equation*}
\Pi (p)=\left( \dfrac{q}{2}+\delta (1-q)\left[ \tau \left( 1-\dfrac{q}{2}%
\right) -\frac{\tau (1-q)}{2}+\widetilde{\tau }(q)(1-F(p))\right] \right) p.
\end{equation*}

Equating these expected profits with the profit of setting a price equal to
the upper bound of the distribution gives the equilibrium price distribution
as 
\begin{equation}
F(p)=1+\eta -\eta \frac{\overline{p}}{p},\ \mbox{with support}\ [\underline{p%
},\overline{p}]  \label{eq:CDF}
\end{equation}%
where 
\begin{equation}
\eta =\frac{\frac{q}{2}+\delta (1-q)\left( \tau \left( 1-\frac{q}{2}\right) -%
\frac{\tau (1-q)}{2}\right) }{\delta (1-q)\widetilde{\tau }(q)}>0,
\label{eta}
\end{equation}%
and $\underline{p}=\frac{\eta }{1+\eta }\overline{p}$ solves $F(\underline{p}%
)=0$, whereas $\overline{p}=\min \left\{ r,v\right\} $. The fraction $\eta $
is the ratio of consumers who do not compare prices to those that do compare
prices (as in the traditional models of \cite{varian1980} and \cite%
{stahl1989}). Here, the fraction of consumers who are informed about only
one price consists of the fraction of searching consumers and those
non-searching consumers who receive only one price quotation from friends,
while the fraction of consumers who are informed about both prices only
consists of non-searching consumers who receive through their social network
information about the prices of both firms.

Let us first discuss the extreme cases where $\delta =0$ or $\delta =1.$ In
these cases, the Diamond paradox emerges and only an equilibrium with no
trade exists.\ If $\delta =0,$ there is no advantage to not search and no
consumer ever compares prices. The left-hand side of (\ref{eq:IC_after})
reduces to $v-E[p]$ and the right-hand side to $s$. Consumers choose to be
active if the expected price is below $v-s$ and drop out of the market if it
is greater than $v-s$. From (\ref{eta}) it is clear that $\eta $ (and thus
expected price) become infinitely large. Hence, consumers prefer not to
search. On the other hand, if $\delta =1,$ there is no advantage to search
so that $q=0.$ If no consumer acquires information herself, no one compares
prices and again $\eta $ (and thus expected price) become infinitely large.
In this case, the ex ante probability that a consumer's friends have
observed two different prices $\widetilde{\tau }(0)=0.$

\begin{proposition}
If $\delta =0$ or $\delta =1,$ there does not exist an equilibrium with
active trade.
\end{proposition}

Having explained the different conditions that should hold in an RPE with
positive sales, we are now able to provide the main result of this section,
namely that for any $0<\delta <1$ an equilibrium exists if the search cost
is sufficiently small.

\begin{theorem}
\label{theorem:RPERG} For any given $0\leq t(1)<1$ and $0<\delta <1$, there
exists a $\overline{s}\leq v$ such that an RPE exists if, and only if, $%
s\leq \overline{s}$. If an RPE exists it is determined by the triple $%
(q,r,F(p))$ solving \eqref{eq:stoppingrule}, \eqref{eq:IC_after}, and %
\eqref{eq:CDF}. Furthermore, as $s\rightarrow 0$ the optimal search
probability $q$ converges to 1, while price dispersion disappears with $%
\overline{p}=\underline{p}=v.$
\end{theorem}

The proof of the proposition is given in the appendix. The main intuition
can be understood as follows. If a fraction $0<q<1$ of consumers actively
search themselves, while the others are passive, then there is a strictly
positive probability that some of the inactive consumers are informed about
two prices so that the price distribution of firms is non-degenerate. The
main challenge then is to find a value of $q$ such that consumers are
indifferent between searching and non-searching. The proof shows that when $%
s\leq \overline{s}$ there exists such a $q$. It is clear that $\overline{s}$
depends on the exogenous parameters and, in particular, in line with\
Proposition 1, that $\overline{s}$ approaches $0$ as $\delta $ approaches $0$
or $1$.

The interesting aspect is that if a fraction $0<q<1$ of consumers search,
while the others do not, then as long as some consumers have more than one
connection, i.e., $t(1)<1,$ it is the fraction of the non-searching
consumers who are free-riding that provide a positive service to the
searching consumers as they are the ones who comapre oth prices. Thus, the
non-searching consumers play a crucial role to resolve the Diamond paradox.

Importantly, if $s$ becomes arbitrarily small, price dispersion must vanish
as consumers can obtain another price quote at virtually no additional cost.
If $s=0,$ any non-negative price $p^{\ast }\leq v$ can be sustained in
equilibrium. Consumers' strategy is to search ($q=1$) and $r=p^{\ast }:$
they buy immedietaly if they observe a price $p\leq p^{\ast }$, otherwise
they continue to search. If $s\rightarrow 0$ the model selects the
equilibrium where firms price at the monopoly level. Given that price
dispersion disappears, the share of non-searching consumers must disappear
as non-searching consumers pay almost the same price as searching consumers,
but incur a cost of waiting as $0<\delta <1$. For any $s>0$ firms have the
market power to raise their price up to the consumers' reservation price,
which is larger than the expected price. As in the Diamond paradox, this
will raise the expected price until the monopoly price $v$. Interestingly,
and different from the Diamond paradox, consumers choose to be search as for
any $s>0$ there remains some price dispersion and the expcted price is
smaller than $v.$

\begin{figure}[h]
\centering
\includegraphics[width=.6\linewidth]{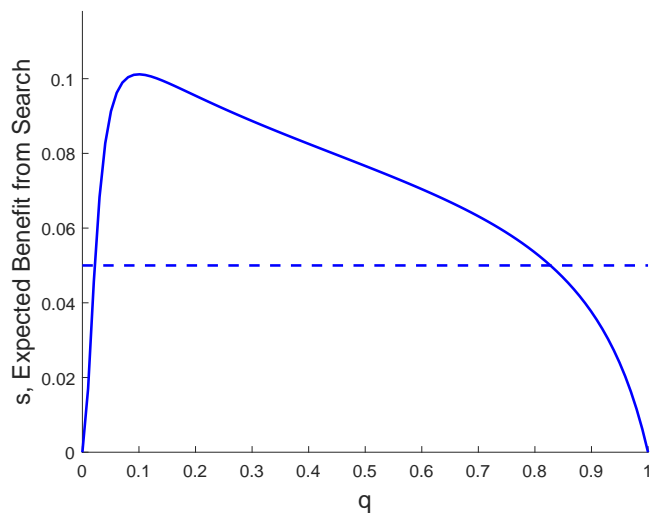} 
\captionsetup{justification=centering}
\caption{An illustration of existence of an RPE for $\protect\delta =0.9$, $%
s=0.05$ and $t(k)=nk^{\protect\gamma }$, where $\protect\gamma =-1$, $%
\overline{k}=100$ and $\sum_{k \in O}nk^{\protect\gamma}=1$.}
\label{fig:after_sym}
\end{figure}

Figure \ref{fig:after_sym} illustrates the equilibrium construction. The
horizontal axis represents the fraction $q$ of searching consumers, while
the cost and the expected benefit of search are presented on the vertical
axis. The solid curve represents the expected benefits, while the dashed
horizontal line represents the cost of search. In the proof we show what the
figure presents, namely that when $q$ approaches 0 or 1 the expected benefit
of search approaches 0. This is quite intuitive: if $q$ approaches 0 or 1,
there are very few consumers who compare prices either because there are
almost no non-searching consumers ($q$ \ close to 1) or because there is
almost no information that is diffused in the system ($q$ \ close to 0).
More formally, this can be seen as $\eta \rightarrow \infty $ if $q$
approaches 0 or 1. Thus, firms exercise their market power, and prices get
close to $v.$ As for interior values of $q$ the expected benefits are
positive and continuous in $q$, it must be the case that for small enough
values of $s$ an equilibrium exists.

The figure shows that for (given) small enough values of $s$ there are two
intersection points and, hence, two equilibrium values of $q$ where the
market is active.\footnote{%
For certain parameter configurations with high search costs, there may exist
four equilibria with positive trade. For our comparative static analysis, we
focus on sufficiently small search costs such that only two equilibria with
active trade exist.} One may argue, however, as in other search models (see,
e.g. \cite{burdettjudd1983}, \cite{fershtmanfishman1992}, \cite%
{janssenmoraga2004} and \cite{honda2015}) that the equilibrium corresponding
to the higher search probability can be called a ``stable" equilibrium in
the sense that if the real search probability falls (slightly) short of the
equilibrium value the expected benefit of search exceeds the cost so that
consumers have an incentive to search more intensively. It follows that if
the search cost asymptotically approaches zero, the optimal search
probability in the active equilibrium approaches 1.\footnote{%
The figure also shows there is a stable equilibrium where the market is
inactive ($q=0$) and (\ref{eq:IC_after}) does not hold.} The comparative
static analysis focuses on this ``stable" equilibrium.

\section{Comparative Statics}

\label{sec:comp_stat}

Given the equilibrium characterization, we now can provide insights into how
market outcomes depend on exogenous parameters. We will first focus on the
impact of network structure on equilibrium prices, before concentrating on
the speed of communication in the network (represented by $\delta $), and
the impact of the cost of searching $s$. A social network like Facebook has
significantly increased the number of connections people have (although
there remain a non-negligible fraction of consumers who do not use Facebook
or other social networks) and the speed of information diffusion through the
network. Online markets have significantly reduced search cost $s$. In this
section, we discuss the implications of these effects on market outcomes,
especially the expected market price. Unless explicitly discussed otherwise,
the changes in firms' profits is perfectly in line with expected price (as
in an RPE eventually all consumers buy). In an RPE, the expected market
price is proportional to $s$ and given by $E[p]=s\left( \frac{\eta \ln
\left( 1+\frac{1}{\eta }\right) }{1-\eta \ln \left( 1+\frac{1}{\eta }\right) 
}\right) ,$ which is increasing in $\eta .$

We investigate the limiting behavior when the network of consumers is dense
and all consumers tend to have many links. One may think that if all
consumers potentially get information from many friends, competition would
prevail and prices converge to marginal cost with price dispersion being
eliminated. Surprisingly, however, as the next result shows, price
dispersion remains an essential feature of any reservation price
equilibrium, and even in the limit when all consumers have infinitely many
links, price dispersion remains bounded away from 0.

\begin{proposition}
\label{prop:t_infty} For any $s>0,$ any RPE with active search is
characterized by price dispersion even if all consumers tend to have
infinitely many links.
\end{proposition}

A consequence of the limiting price dispersion is that prices do not
converge to marginal cost. The reason that price dispersion remains is that
in any RPE without price dispersion, the reservation price, and hence, the
expected price should be infinitely large. But then consumers are better off
not searching.

It is important to understand that for this result to hold the first search
should be costly. If this were not the case, all consumers would search and
the probability of obtaining the second price from friends then approaches
one as the number of links to each consumer in the population becomes large.
As a consequence, market frictions would disappear and prices converge to
marginal cost. This also explains why in \cite{galeotti2010} prices do
converge to marginal cost if the network of consumers gets dense: it is the
consequence of the first search being assumed to be free in his model.

\begin{figure}[!htb]
\centering
\captionsetup{justification=centering} \minipage{0.48\textwidth} 
\includegraphics[width=\linewidth]{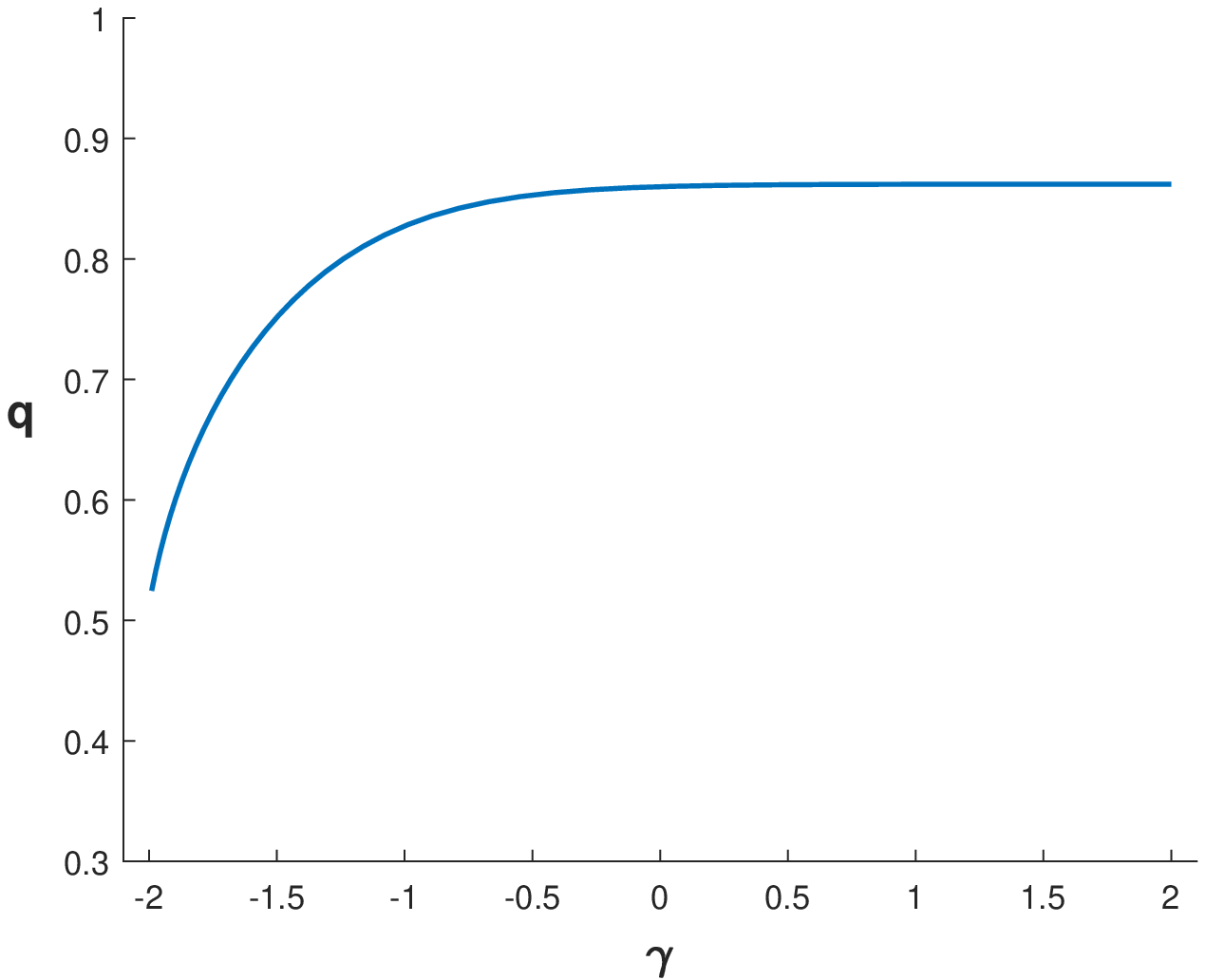} 
	\caption{Impact of network density on the share of active consumers}
	\label{fig:cs_t_q}
\endminipage
\hfill \minipage{0.48\textwidth} 
\includegraphics[width=\linewidth]{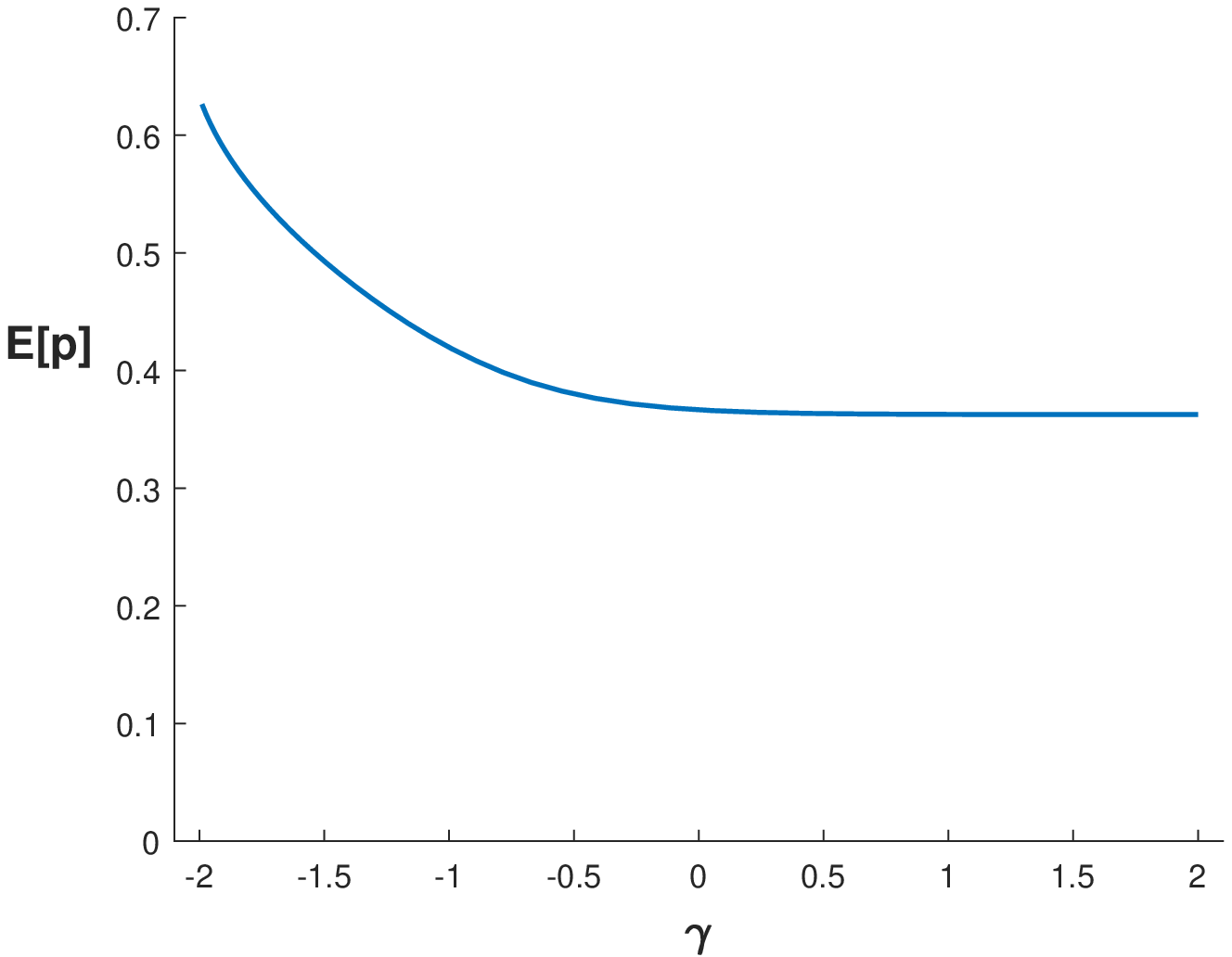}    
	\caption{Impact of network density on expected price}
	\label{fig:cs_t_Ep}
	\endminipage
\end{figure}

By means of numerical simulations, we can analyze intermediate cases of
network connectivity. We assume the social network can be described by a
random graph that follows a power law\footnote{%
Empirical analysis has demonstrated that many social networks can be
described by a power-law of the form we assume here (see, for example, \cite%
{price1965}).} $t(k)=nk^{\gamma }$, where larger $\gamma $ values stand for
denser networks with consumers having more connections. In particular,
comparing two networks with probabilities of connections being given by $%
t(k) $ and $\widetilde{t}(k),$ the network with probabilities $t(k)$ is
generated by a higher $\gamma $ (and thus, denser) if there exists a $%
\widetilde{k}\in O$ such that $t(k)<\widetilde{t}(k)$ for all $k<\widetilde{k%
},$ whereas $t(k)>\widetilde{t}(k)$ for all $k>\widetilde{k}.$ In Figures %
\ref{fig:cs_t_q} and \ref{fig:cs_t_Ep},\footnote{%
The following parameter values have been used: $v=1$, $s=0.05$, $\delta =0.9$%
, and $\overline{k}=100$.} we gradually increase $\gamma $ from $-2$ to $2$.

Figure \ref{fig:cs_t_Ep} depicts the impact of network density on the
expected price. It shows that the expected price is decreasing in $\gamma $
and that certainly when the network is not very dense ($\gamma $ is small),
this impact is quite strong as the expected price may decrease by around $%
50\%$ as $\gamma $ increases from around $-2$ to $0.$ The main, direct
impact can be understood by noting that if the network is described by the
power law $t(k)=nk^{\gamma }$ it follows from\ (\ref{eta}) that the ratio $%
\eta $ is given by 
\begin{equation*}
\eta =\frac{\dfrac{q}{2}+\delta (1-q)\sum_{k\in O}nk^{\gamma }\left[ \left(
1-\dfrac{q}{2}\right) ^{k}-\dfrac{\left( 1-q\right) ^{k}}{2}\right] }{\delta
(1-q)\left( 1+\sum_{k\in O}nk^{\gamma }\left[ \left( 1-q\right) ^{k}-2\left(
1-\dfrac{q}{2}\right) ^{k}\right] \right) }.
\end{equation*}%
As the term in square brackets in the numerator is positive and decreasing
in $k$, while the term in square brackets in the denominator is negative and
increasing in $k$, it follows that networks with higher $\gamma $'s put
relatively more weight on a higher number of connections, which implies that 
$\eta $ is decreasing in $\gamma .$ As lower $\eta $'s reflect the fact that
there are relatively more price comparing consumers in the market, the
direct effect puts downward pressure on prices. There is, however, also an
indirect effect through the consumers' search probability $q$. A decrease in
price levels is associated with a decrease in price dispersion, as measured
by the difference $E[p]-E_{\min }[p]$. This makes it more attractive for
consumers to search as the main benefit of not searching, namely being
informed about both prices and therefore being able to buy at the lowest of
the prices, becomes smaller. In addition, as the expected price is lower,
consumers would like to have the larger benefit of a purchase now rather
than having to wait for information through friends. The associated increase
in the share of searching consumers as illustrated in Figure \ref{fig:cs_t_q}
increases $\eta $ and therefore also increases expected price. This indirect
effect is, however, smaller than the direct effect and thus, the overall
impact on expected price is decreasing.

Note that the above result on the fraction of consumers actively acquiring
information is strikingly different from \cite{galeotti2010} and \cite%
{galeottigoyal2010} where an agent's probability of actively acquiring
information negatively correlates with the number of links she has.
Intuitively, one might expect that the more connections a consumer has the
less likely she is to become active as, all else being equal, the
probability that she obtains information from friends rises. In our case,
however, it is the combined price effect through a lower expected price and
lower price dispersion as measured by $E[p]-E_{\min }[p]$ that overrides
this intuitive effect.

\begin{figure}[!htb]
\centering
\captionsetup{justification=centering} \minipage{0.48\textwidth} 
\includegraphics[width=\linewidth]{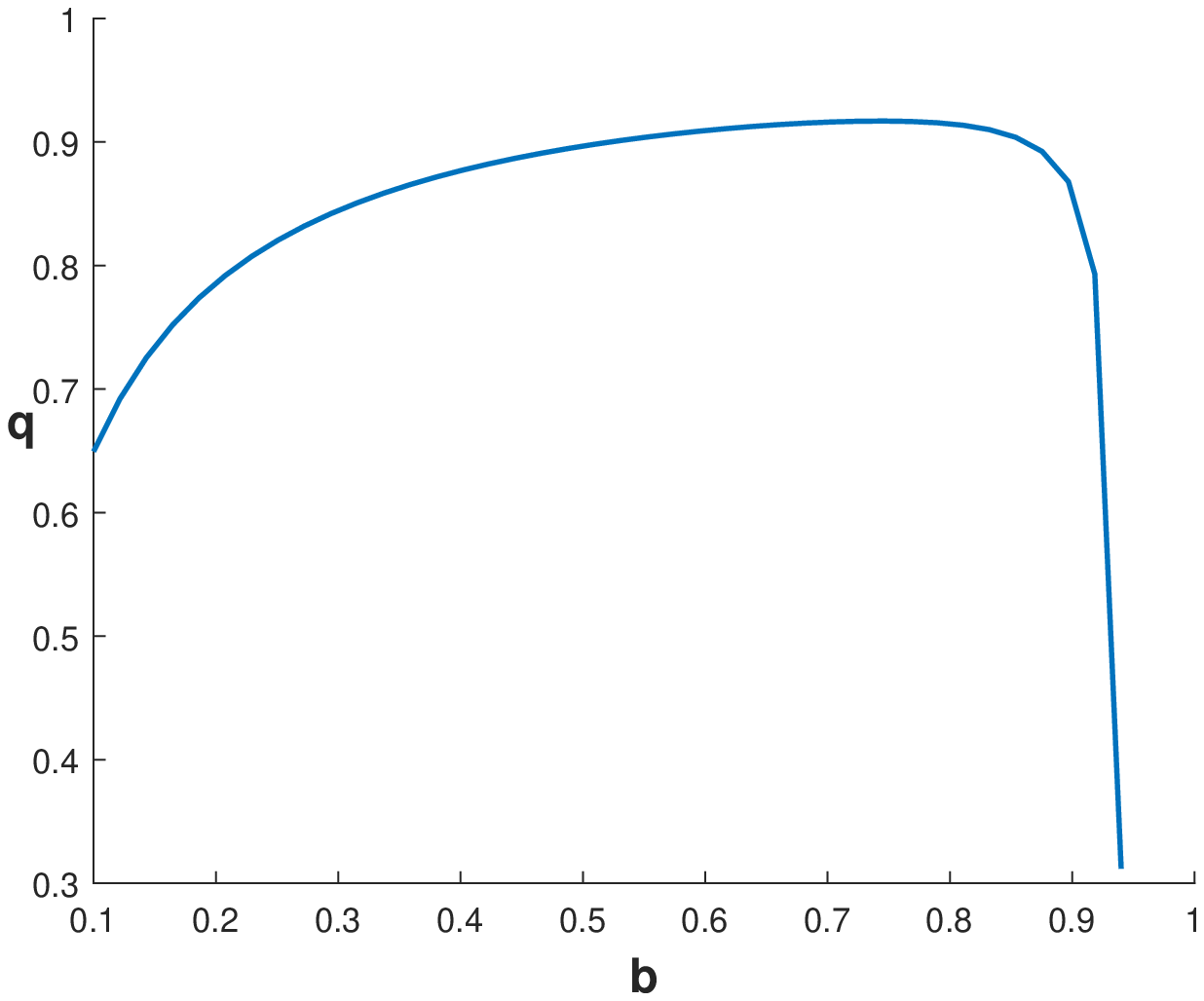}    
\caption{Impact of $\protect\delta$ on the share of active consumers}
\label{fig:delta_q}
\endminipage
\hfill \minipage{0.48\textwidth} 
\includegraphics[width=\linewidth]{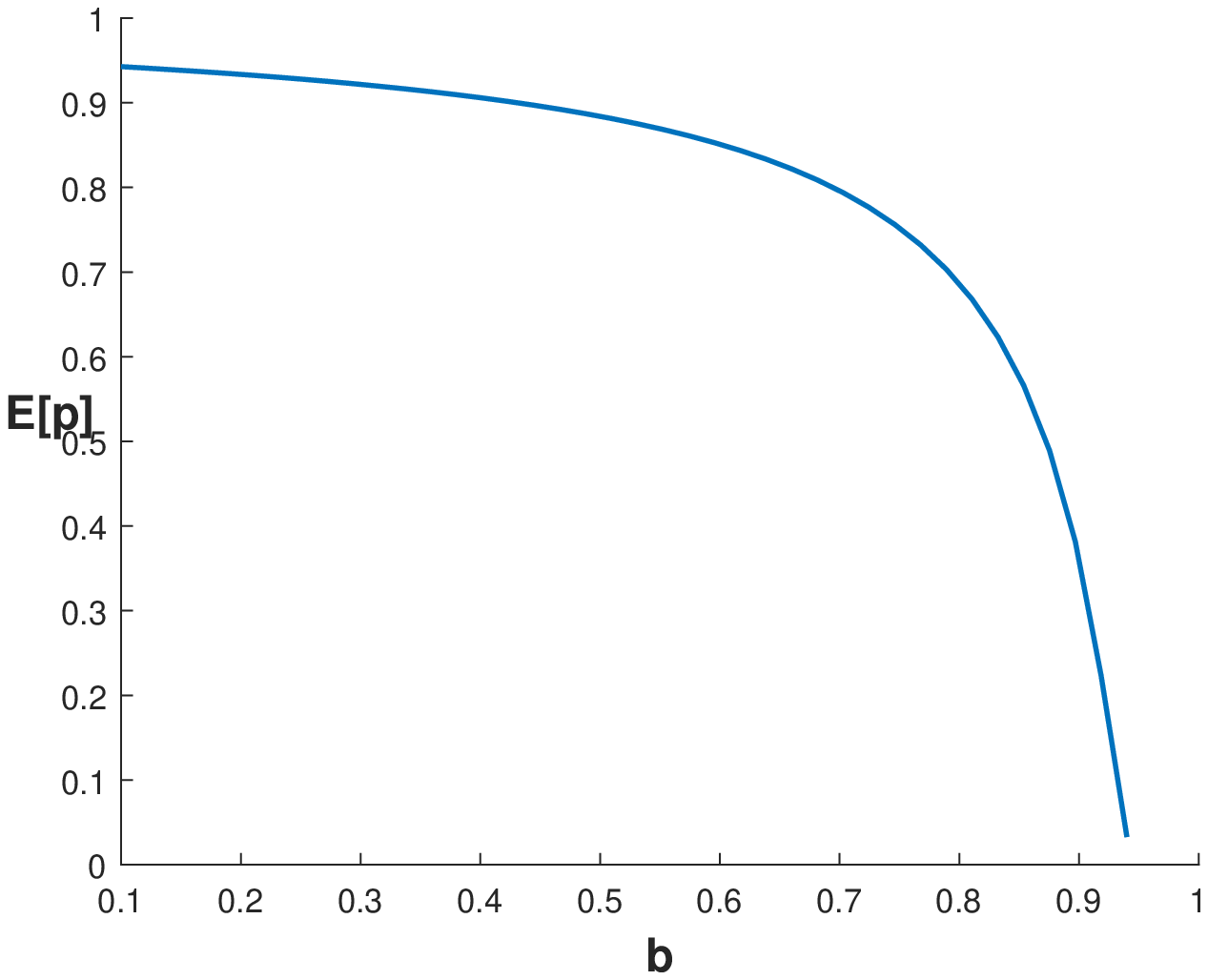}    
	\caption{Impact of $\protect\delta$ on expected price \newline
	}
	\label{fig:delta_Ep}
\endminipage
\end{figure}

Next, we investigate the impact of a change in $\delta $. Social media have
significantly increased the speed with which consumers may share
information, which in our model is measured by $\delta $. We know that a
higher $\delta $, resulting in faster information transmission, permits a
consumer to access information quickly from friends, but that this may have
repercussions on the incentives to search. Proposition 3 establishes that at
extreme values of $\delta $ an equilibrium with positive trade does not
exist. Numerical simulations in Figures \ref{fig:delta_q} and \ref%
{fig:delta_Ep} show the effect of an increase in $\delta $ on the share of
searching consumers and the expected price when we change the value of $%
\delta $ in an intermediate range (here from $0.1$ to $0.94$)$.$ For larger
values of $\delta $, an equilibrium with search ceases to exist.\footnote{%
The Figures are drawn for the following parameter values: $v=1,$ $s=0.05$, $%
t(k)=nk^{\gamma }$, $\overline{k}=100$, and $\gamma =0$.} As before, there
is a direct and an indirect effect of $\delta $ on prices. The direct effect
can be seen by taking the derivative of $\eta $ with respect to $\delta $ in
(\ref{eta}). It is easy to see that this derivative is negative so that an
increase in the speed of information processing in the population increases
the share of price comparing consumers. This effect reduces prices. There is
also an indirect effect via $q,$ however. For small values of $\delta $ the
indirect effect is very similar to the indirect effect we mentioned in
relation to the impact of $\gamma $, namely that as prices and price
dispersion decline, consumers have more incentives to become active
searchers themselves. In this case, there is, however, also a direct effect
of $\delta $ on $q$ (as shown in Equation (\ref{eq:IC_after})). As $\delta $
becomes large, there is almost no downside to waiting anymore and consumers
massively change their behavior and do not search. As, focusing on stable
equilibria, $\eta $ is increasing in $q$ this further strengthens the direct
effect so that the total effect of $\delta $ on the expected price becomes
more pronounced.

\begin{figure}[!htb]
\centering
\captionsetup{justification=centering} \minipage{.48\textwidth} 
\includegraphics[width=\linewidth]{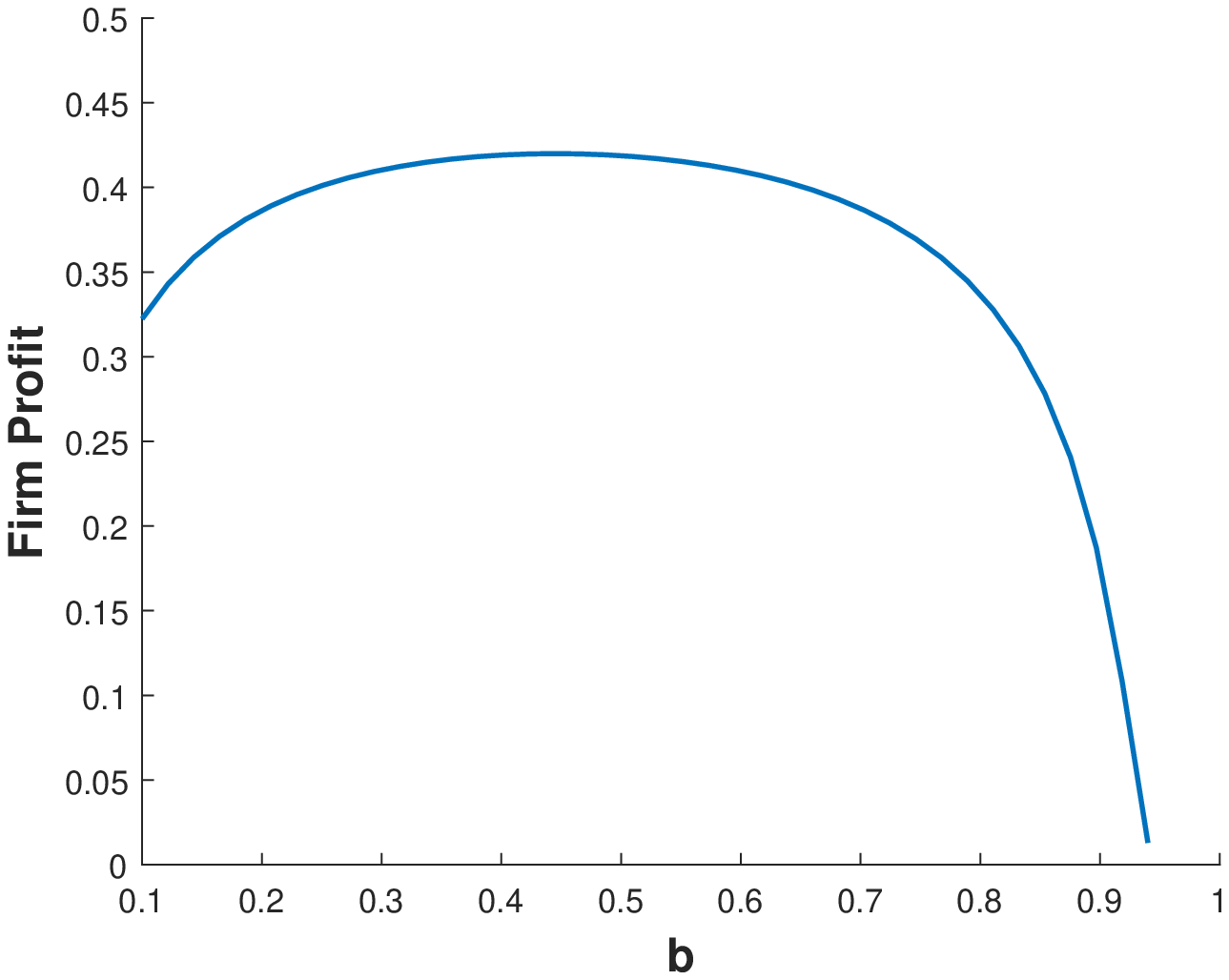}    
	\caption{Impact of $\protect\delta$ on firm profit}
	\label{fig:delta_Pi}
\endminipage
\hfill \minipage{.48\textwidth} 
\includegraphics[width=\linewidth]{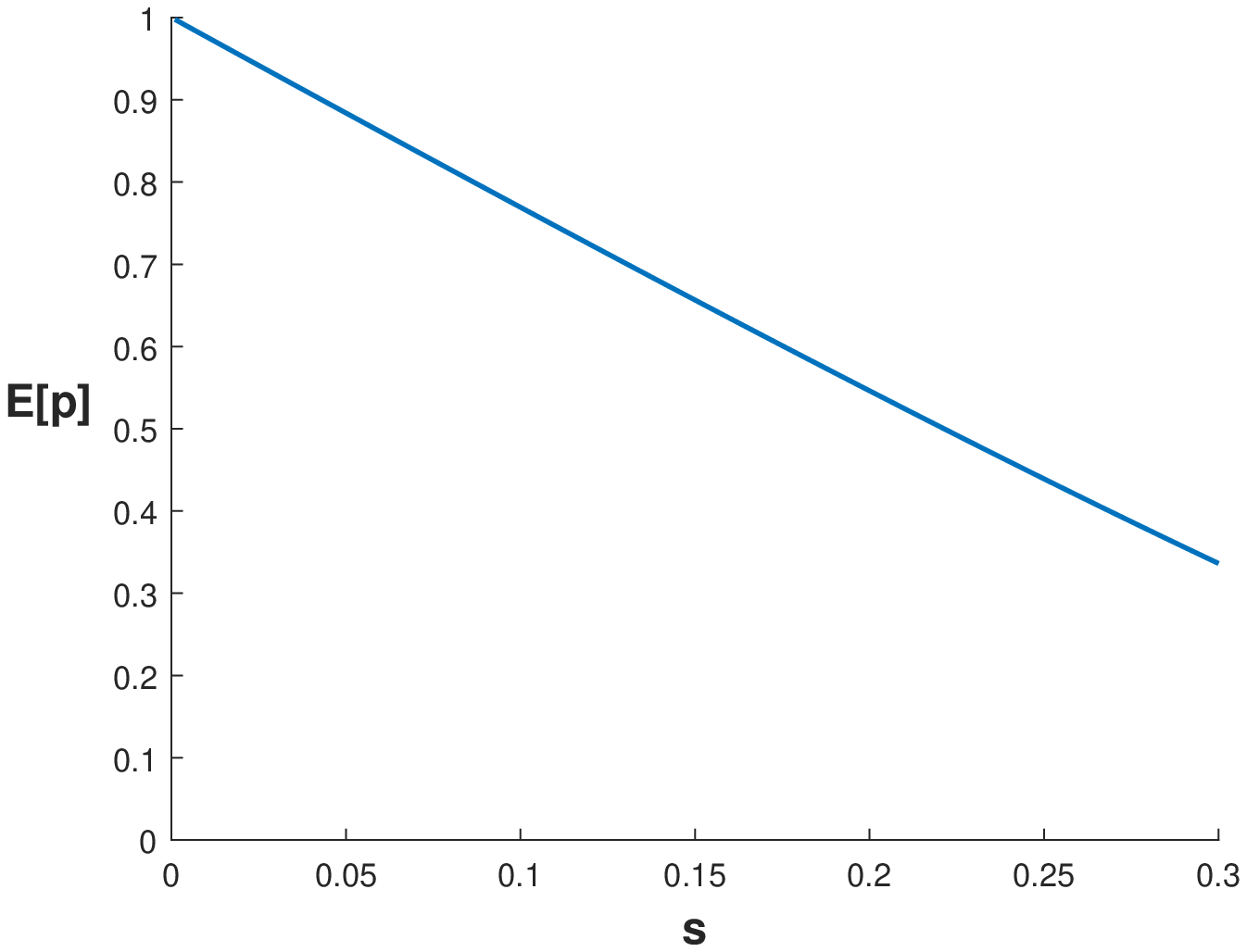}    
	\caption{Impact of $s$ on expected price}
	\label{fig:s_Ep}
\endminipage
\end{figure}

Interestingly, the speed of information diffusion also has a non-monotonic
effect on firms' profits as shown in Figure \ref{fig:delta_Pi}. When the
speed of information diffusion is relatively low, firms have an incentive to
increase the number of searching consumers as these consumers buy
immediately, whereas the sales to non-searching consumers are heavily
discounted.\ Thus, even though expected price is monotonically decreasing in
the speed of information diffusion, the fact that the fraction of searching
consumers is increasing offsets the decrease in revenue per consumer. On the
other hand, when the speed of information diffusion is already relatively
high to begin with, firms would not want to increase it further. At some
point, if the speed of information diffusion is very large, both firms and
consumers suffer from a further increase in the diffusion speed as an
equilibrium with active trade ceases to exist.

Finally, we proceed by studying the impact of a change of $s$. We already
know that the optimal search probability goes to 1 in the limit when $s$
goes to zero, suggesting that as $s$ starts increasing from $0$ the optimal
search probability decreases. The next proposition shows that the underlying
effects also hold true outside the region where $s$ is close to $0$.

\begin{proposition}
\label{prop:cs_bs} In any stable RPE, the share of active consumers is
decreasing in $s$.
\end{proposition}

We know from Theorem 1 that in the limit when $s$ is arbitrarily small,
prices converge to the monopoly price $v$. This immediately implies that for
small enough search cost, the expected price must be decreasing in $s$ as
more consumers make price comparisons and firms will compete for these
consumers. As $\eta $ is increasing in $q$ and expected price is increasing
in $\eta ,$ it follows that the expected price is decreasing in $s$ for
larger values of $s$ as well, as illustrated in Figure \ref{fig:s_Ep}.%
\footnote{%
We use the following parameter values for the figure: $v=1$, $%
t(k)=nk^{\gamma }$, $\overline{k}=100$, $\gamma =0,\delta =0.5$.}

\section{A Foundation for the Baseline Model}

There are many ways to model the interaction between sequential search and
word-of-mouth communication. The baseline model we have analyzed so far is
simplistic as (\textit{i}) it only allows for two periods and (\textit{ii})
a searching consumer cannot decide to not continue to search and wait for
information from friends. In\ this section we show that for small search
costs the main results of the previous sections continue to hold in a more
general multi-period model where in every period consumers may decide to buy
using the information they have, search themselves, or not search (and go to
the next period where they may receive more information from friends)
provided that $t(1)>0.$

We also show that the formal equivalence between the equilibrium outcome of
the baseline model and the more general model depends on consumers only
sharing information they have acquired themselves. This is similar to the
setting studied in \cite{bramoullekranton2007}, \cite{banerjidutta2009} and 
\cite{ellisonfudenberg1995} where information decays after one link in the
network. The qualitative results continue to hold, however, if information
flows to all directly or indirectly connected consumers.

So, let us first analyze the general model where at every decision moment
consumers can choose to search themselves, buy from one of the firms from
which they have obtained information or not search with the following three
conditions: (\textit{i}) a consumer can only buy if she is informed of at
least one price, (\textit{ii}) if a consumer decides to not search, then she
``waits'' and she is moved to the next period where her pay-offs are
discounted by $\delta ,$ and (\textit{iii}) information exchange takes place
at the beginning of every period. Thus, the following dynamic unfolds. After
firms have chosen their prices and the network is formed, consumers at the
beginning of their first ``search period" can choose to search or not
search. Not searching consumers have to wait for taking actions until the
next period. After their first search, searching consumers can decide to
buy, continue to search (and then decide where to buy) or not search anymore
in this round. Then, the second ``search period" starts and information is
exchanged: everyone that acquired information shares with her immediate
friends, and all consumers that did not buy yet can choose to search, buy or
not search subject to the above three conditions. This process continues as
long as some consumers have not bought yet and still consider buying at some
point.

We now show by means of a series of lemmas that in this general model, (%
\textit{i}) all consumers who decided not to search in the first ``search
period" and who did not receive any price information from friends at the
beginning of the second ``search period" will search themselves and (\textit{%
ii}) all consumers who have acquired price information either from friends
or through their own search activity will buy immediately, \textit{i.e.},
they will not continue to search and they will not decide not to search and
wait until the next period.\ It then follows that the equilibrium outcome is
exactly the same as in the baseline model so that the fraction $q$ of
searching consumers in the first ``search period" is determined by the same
indifference condition (\ref{eq:IC_after}) and also $\eta $ is determined by
(\ref{eta}).\ 

First, as for any $0<q<1$ some consumers will be informed about both prices,
there cannot be mass points in the symmetric price distribution as it will
pay for firms to undercut. But then a firm that charges an upper bound $%
\overline{p}>r$ will not sell to any consumer.

\begin{lemma}
\label{lem:noatom_gen} $F(p)$ does not have atoms and $F(r)=1$.
\end{lemma}

Next, we show that for $s$ close enough to 0, it should be that $q$ is close
to 1. As the reservation price $r$ continues to be defined by \ref%
{eq:stoppingrule}, it follows that for $s$ close to 0 $r\approx E[p],$ but
given that there are no mass points, it then also follows that $E[p]\approx
E[\min \left\{ p_{1},p_{2}\right\} ].$ As price dispersion only disappears
for $q$ close to $0$ or $1$, the stable equilibrium is the one where $q$ is
close to 1.

\begin{lemma}
If $s$ is close enough to 0, then $q$ is close to 1.
\end{lemma}

Another point in the overall argument is that consumers who did not search
in the first ``search period" want to search in the second ``search period"
if they did not get information from friends. The argument essentially is
that it is much more likely to be informed by friends in the beginning of
the second period than it is in the beginning of the third period. What is
important in this regard is that the probability that a consumer will be
informed by a friend at the beginning of the second period is given by $\sum
t(k)\left[ 1-(1-q)^{k}\right] ,$ while the probability $x$ that a consumer's
friend will be informed by their friends, knowing that I did not search,
equals $x=\sum t(k)\left[ 1-(1-q)^{k-1}\right] .$ This latter probability is
important as, in case information decays after one link in the network, a
consumer cannot count on receiving information from friends in future
periods if they bought the product after receiving information from their
friends that they do not pass on. The probability that I will be informed by
my friends in the third period after I did not receive information from them
in the second period is then maximally equal to $\sum t(k)(1-x)^{k}$ as at
that moment I can only be informed by friends who searched themselves in the
second period (and who thus did not get information from their friends)$.$
For $s$ close to 0 and $q$ close to 1, $x\approx 1-t(1)$ so that $\sum
t(k)(1-x)^{k}\approx \sum t(k)t(1)^{k}<t(1)<1.$ Thus,

\begin{lemma}
If $s$ is close enough to 0, then consumers who did not search from the
start want to search after they did not get information from friends.
\end{lemma}

It remains to be shown that if a consumer is informed about any price $p<r,$
she does not want to wait for the next period. To this end, denote by $\rho $
the smallest price at which a consumer who is informed about that price
would be indifferent between not searching (and wait until the next period)
and buying. An active consumer who waits after searching the first firm
receives a price quote of the other firm from friends with probability $%
1-\tau (1-\frac{q}{2})$. Thus, we can write the payoff of an active buyer
who observes price $\widetilde{p}$ and waits as $\delta \left[ v-\left(
1-\tau (1-\frac{q}{2})\right) F(\widetilde{p})E[p|p<\widetilde{p}]-\left[
1-\left( 1-\tau (1-\frac{q}{2})\right) F(\widetilde{p})\right] \widetilde{p}%
\right] $. Then, since the payoff from buying at price $\widetilde{p}$ is $v-%
\widetilde{p}$, $\rho $ is implicitly given by 
\begin{equation}
(1-\delta )\left( v-\rho \right) =\delta \left( 1-\tau (1-\frac{q}{2}%
)\right) F(\rho )\left( \rho -E[p|p<\rho ]\right) ,  \label{eq:rho}
\end{equation}

The proof that a consumer does not want to wait if she is informed about a
price $p<r,$ is in two parts. First, in the Appendix we show that the RHS of
(\ref{eq:rho}) is increasing in $\rho $ so that

\begin{lemma}
\label{lem:uniquerho_gen} If there is some $\rho <r$, it must be unique.
\end{lemma}

The last part of the argument builds on the fact that equilibria with active
markets only exist if at least some consumers are searching and for this it
must be true that consumers, who are \textit{ex ante} identical, at least
weakly prefer to search than not to search. We show that this weak
preference implies that consumers would strictly prefer to continue to
search than not to search after having observed a first price quote. The
only reason for a searching consumer to wait after having observed a first
price quote is that she hopes to be able to economize on the search cost $s$
by getting informed about the other price quote via a friend. However, the
chance of being informed through the social network about the second price
that is not yet observed is smaller than the chance of being informed about
any price. Thus, if a consumer weakly prefers to search before observing a
price, she strictly prefers to search than to wait after observing a price.\ 

\begin{lemma}
\label{prop:r<rho} In any equilibrium where there is active search, we have
that $r\leq \rho $.
\end{lemma}

Together with $F(r)=1$ it follows that in any equilibrium with trade
consumers immediately buy after observing a price. Thus, we have the
following result:

\begin{proposition}
If $s$ is small enough and information decays after one link in the network,
then all equilibria of the general model coincide with the equilibria of the
baseline model.
\end{proposition}

\subsection{Information spreading through the network}

So far in this section we have continued to assume that consumers can only
spread information if they have searched for it themselves, i.e.,
information decays after one link. In this subsection we consider an
alternative case where information does not decay. The most straightforward
way for information not to decay is that at the beginning of round 2 all
information is spread to everyone who has a direct or indirect link to each
other.\footnote{%
Alternatively, one could think that every period information is spread to
direct friends only so that it takes two periods to obtain information that
friends of friends acquired, and so on. As this is somewhat more cumbersome
to analyze, we restrict ourselves to all information becoming available to
everyone immediately.}

It is known that as the number of agents in a network grows without bounds,
the network becomes fully connected if the probability that two random
consumers are linked is larger than some cutoff probability. For example, in
networks where the distribution of links over some $n$ number of agents is
Poisson distributed, this cutoff probability is $\ln (n)/n$ (Theorem 4.1. in 
\cite{jackson2008})$.\ $Thus, the network becomes almost surely connected as 
$n\rightarrow \infty $ if the probability that a buyer has a link is greater
than $\ln (n)/n$. In this section we assume that the network structure of
our model satisfies the above condition.

It is clear that $F(r)=1$ should hold again. If all information is spread
instantaneously to all consumers that are directly or indirectly linked to
each other, a non-searching consumer can buy almost surely at the lowest of
both prices so that the indifference condition determining $q$ is equal to%
\begin{equation}
v-E[p]-s=\delta (v-E_{\min }[p]),  \label{ind section 5}
\end{equation}%
while firms' profit is given by 
\begin{equation*}
\Pi (p)=\left( \dfrac{q}{2}+\delta (1-q)(1-F(p))\right) p.
\end{equation*}%
It follows that 
\begin{equation}
F(p)=1-\frac{q}{2\delta (1-q)}\frac{r-p}{p}.  \label{F(p) section 5}
\end{equation}%
Thus, we have a similar result as in\ Theorem 1:

\begin{proposition}
\label{prop:information flows}For any $0<\delta <1$, there exists a $%
\overline{s}\leq v$ such that an RPE exists if, and only if, $s\leq 
\overline{s}$. If an RPE exists it is determined by the triple $(q,r,F(p))$
solving \eqref{eq:stoppingrule}, \eqref{ind section 5} and 
\eqref{F(p) 
section 5}. Furthermore, as $s\rightarrow 0$ the optimal search probability $%
q$ converges to 1, while price dispersion disappears with $\overline{p}=%
\underline{p}=v.$
\end{proposition}

\section{Information Asymmetry about Connections}

If consumers know the number of links they have before engaging in search,
not all consumers search with the same probability. There may be consumers
whose optimal search probability is between zero and one, and there will be
ones who either definitely search or do not search at all. As consumers with
more connections are more likely to obtain information via their social
network than consumers with fewer links, the expected payoff from waiting is
increasing in the number of connections. Thus, if a consumer with, say, $%
\widehat{k}$ connections is indifferent between becoming an active searcher
and staying passive, so that $0<q(\widehat{k})<1,$ then all consumers with
more connections wait to obtain information through their social network,
while those with fewer connections definitely search. If there is no
consumer who is indifferent, then we can set $\widehat{k}$ to be the largest
number such that $q(\widehat{k})=1.$ Clearly, for any $s>0$ it cannot be the
case that $\widehat{k}=\overline{k}$ and $q(\overline{k})=1$ as then all
consumers would search themselves and no one would compare prices.

\begin{proposition}
\label{prop:cutoff_k} Let $\widehat{k},1\leq \widehat{k}\leq \overline{k}$
be as defined above. Consumers with a number of friends less than $\widehat{%
k }$ search with probability one and consumers with more than $\widehat{k}$
friends do not search.
\end{proposition}

Thus, consumers with more connections are more inclined to not search
themselves and wait for information from their friends. It is clear that
from a social welfare perspective, the wrong set of people search. Keeping
the fraction of active consumers constant, if people with more links would
search instead, there are two effects. First, the welfare loss of passive
consumers who do not acquire any information and still need to search
themselves is smaller as their fraction will be smaller. Second, as more
consumers would make price comparisons, there is a redistribution of surplus
from firms to consumers.

Given this result, the consumer behavior in any equilibrium can be
characterized by three parameters; $\widehat{k},$ $q(\widehat{k})$ and a
reservation price $r.$ In what follows we use $q$ as a short-hand notation
for $q(\widehat{k}).$

Correctly anticipating the optimal behavior of consumers, an individual
firm's expected profit of setting a price equal to $p\leq r$ is equal to 
\begin{equation}
\begin{split}
\Pi (p)=& \frac{1}{2}\widehat{w}p+\delta t(\widehat{k})(1-q)\left[
\sum_{m=1}^{\widehat{k}}\binom{\widehat{k}}{m}w^{m}\left( 1-w\right) ^{%
\widehat{k}-m}\left( \frac{1}{2^{m}}\right) +\frac{(1-w)^{\widehat{k}}}{2}%
\right] p \\
& +\delta t(\widehat{k})(1-q)\sum_{m=1}^{\widehat{k}}\binom{\widehat{k}}{m}%
w^{m}\left( 1-w\right) ^{\widehat{k}-m}\left( 1-\frac{1}{2^{m-1}}\right)
(1-F(p))p \\
& +\delta \left[ \sum_{k=\widehat{k}+1}^{\overline{k}}t(k)\sum_{m=1}^{k}%
\binom{k}{m}w^{m}\left( 1-w\right) ^{k-m}\left( \frac{1}{2^{m}}\right) +%
\frac{(1-w)^{\widehat{k}}}{2}\right] p \\
& +\delta \sum_{k=\widehat{k}+1}^{\overline{k}}t(k)\sum_{m=1}^{k}\binom{k}{m}%
w^{m}\left( 1-w\right) ^{k-m}\left( 1-\frac{1}{2^{m-1}}\right) (1-F(p))p,
\end{split}
\label{eq:asym_profit_after_p}
\end{equation}%
where $\widehat{w}=\sum_{k=1}^{\widehat{k}-1}t(k)+t(\widehat{k})q$ is the
average search probability of a consumer and $w=\frac{\sum_{k=1}^{\widehat{k}%
-1}t(k)k+t(\widehat{k})\widehat{k}q}{\sum_{k=1}^{\overline{k}}t(k)k}$ is
approximately the search probability of a consumer's neighbor (not knowing
how many friends the neighbor has; cf., Section 4.2. in \cite{jackson2008}).

This expression can be understood as follows. A fraction of $\sum_{k=1}^{%
\widehat{k}-1}t(k)$ consumers has less than $\widehat{k}$ links. They search
themselves and visit the firm with probability $0.5$. Since they do not
compare prices they pay the price charged by the firm (if this is not larger
than their reservation price). Consumers with $\widehat{k}$ links, who make
a share of $t(\widehat{k})$ of the population, search with probability $q$
and visit the firm in half of the cases, and these consumers along with
those with fewer links give the first term in \eqref{eq:asym_profit_after_p}%
. With probability $(1-q)$, consumers with $\widehat{k}$ links do not search
and can be informed by their friends about price(s). Each of their friends
searches with an expected probability $w$. Then, consumers obtain
information only about the firm's price if all of their searching friends,
represented by $m$, happen to visit that firm, which occurs with probability 
$\frac{1}{2^{m}}\binom{\widehat{k}}{m}w^{m}\left( 1-w\right) ^{\widehat{k}%
-m} $. If none of her friends search, meaning $m=0$, the consumer searches
herself later and visits the firm half of the time. This gives the second
term. The third term represents the probability that a consumer does not
search and obtains information about both prices from her $m$ searching
friends. The probability of this event is equal to $\left( 1-\frac{1}{2^{m-1}%
}\right) \binom{\widehat{k}}{m}w^{m}\left( 1-w\right) ^{\widehat{k}-m}$. In
this case, a consumer buys from the firm if the other firm charges a higher
price than $p$, which happens with probability $1-F(p)$. Finally, the last
two terms in the profit function account for the share of the population
with more than $\widehat{k}$ links. These expressions are similar to the
ones for the consumers with $\widehat{k}$ links who do not search
themselves. As all waiting consumers buy with a delay (after the information
about prices arrives to them), the payoff from these consumers is discounted
by $\delta $.

The expected profit in \eqref{eq:asym_profit_after_p} can be simplified as 
\footnote{%
Unfortunately, we cannot use probability generating functions to simplify
this expression further due to the fact that consumers search differently
depending on the number of their connections.} 
\begin{equation*}
\Resize{}{ \begin{split} \Pi (p)=& \left[ \frac{\widehat{w}}{2}+ \delta
t(\widehat{k})(1-q)\left( \left(1-\frac{w}{2}\right)^{\widehat{k}}-
\frac{(1- w)^{\widehat{k}}}{2}\right) + \delta
\sum_{k=\widehat{k}+1}^{\overline{k} }t(k)\left( \left(1-\frac{w}{2}\right)
^{k} - \frac{ (1-w)^{k}}{2}\right)\right] p \\ & +\delta\left[
t(\widehat{k})(1-q)\left( 1+(1-w)^{\widehat{k}}-2\left( 1-\frac{w}{2}\right)
^{\widehat{k}}\right) +\sum_{k=\widehat{k} +1}^{\overline{k}}t(k)\left(
1+(1-w)^{k}-2\left( 1-\frac{\widehat{w }}{2}\right) ^{k}\right) \right]
(1-F(p))p, \end{split} }
\end{equation*}%
where, in the square brackets in the first line, we have the share of
consumers who do not compare prices and in the square brackets in the second
line, we have the share of consumers who compare prices.

Equating this expression to the profit that the firm expects to make by
charging $\overline{p}$, we can derive the equilibrium pricing distribution
function: 
\begin{equation}
F(p)=1+\widehat{\eta }-\widehat{\eta }\frac{\overline{p}}{p},\ 
\mbox{with
	support}\ \left[ \underline{p},\overline{p}\right] ,
\label{eq:CDF_after_asym}
\end{equation}%
where (similar to the previous section) $\widehat{\eta }$ is the fraction of
the share of consumers who do not compare prices to the share of consumers
who compare prices. As before, the upper bound of the distribution is not
larger than the reservation price $r$ and $v$, where the reservation price
of a consumer is determined in \eqref{eq:stoppingrule} and $F(p)$ is given
by \eqref{eq:CDF_after_asym}.

Obviously, a full equilibrium analysis under asymmetric information on
network structure is somewhat tedious. For a given $\widehat{k}$ and a $q( 
\widehat{k})\ $one can derive the equilibrium price distribution, but given
the equilibrium price distribution one should check whether the postulated
behavior of consumers is indeed optimal and all consumers with $k>\widehat{
k }$ prefer to be passive and all consumers with $k<\widehat{k}$ prefer to
be active.

The proposition below states the main result for small enough values of $s:$

\begin{proposition}
\label{prop:RPE_after_asym} For any given $0\leq t(1)<1$ and $0<\delta <1$,
and sufficiently small search cost $s$, there exists an RPE given by the
triple $(q,r,F(p))$ and the cutoff $\widehat{k}$, which are determined by %
\eqref{eq:stoppingrule}, \eqref{eq:CDF_after_asym}, and 
\begin{equation*}
\Resize{}{ \dfrac{s}{v} \leq \dfrac{1-\delta}{1 -\delta (1-w)^{\widehat{
k}}+ \frac{\widehat{\eta}}{1-\widehat{\eta} \ln \left(1 +
\frac{1}{\widehat{\eta}}\right)}\left[\delta\left( 1+(1-w)^{\widehat{k}
}-2\left( 1-\frac{w}{2}\right) ^{\widehat{k}}\right) \left(
(1+2\widehat{\eta})\ln \left( \frac{1+\widehat{\eta }}{\widehat{\eta }}
\right)-2\right) + (1-\delta)\ln \left( \frac{1+\widehat{\eta
}}{\widehat{\eta }} \right)\right] }, }
\end{equation*}%
where the inequality holds with equality if $0<q(\widehat{k})<1$.
Furthermore, as $s\rightarrow 0$, the cutoff number of links $\widehat{k}$
equals $\overline{k}$ and the optimal search probability $q$ converges to 1,
while price dispersion disappears with $\overline{p}=\underline{p}=v.$
\end{proposition}

\begin{figure}[h]
\centering
\captionsetup{justification=centering} 
\includegraphics[scale=0.65]{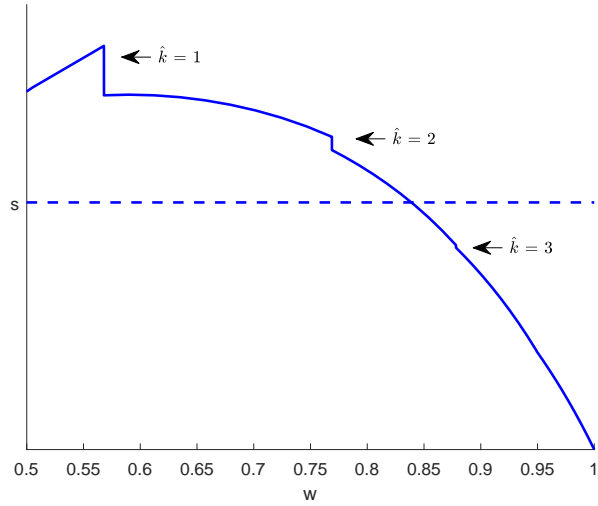}
\caption{Illustration of existence of an RPE when consumers observe the
number of direct links}
\label{fig:after_asym}
\end{figure}

Figure \ref{fig:after_asym} provides an illustrative example of the
Proposition.\footnote{%
The Figure is drawn for the following parameter values: $v=1$, $s=0.025$, $%
\delta =0.92$, $\overline{k}=5$, and $\gamma =-2.5$.} Note that on the
horizontal axis we represent $w,$ the probability a random neighbor is
active. The solid line representing the expected benefit from search has
vertical elements. This is a consequence of the fact that for certain ranges
of search costs, there is no indifferent consumer so that $w,$ and the
underlying decision of consumers whether or not to become active, does not
change. This can be explained as follows. Suppose that the search cost is
such that consumers with $\widehat{k}<\overline{k}$ links search with
strictly positive probability, $0<q(\widehat{k})<1$. This is a situation
represented by one of the non-vertical parts of the stable, right-hand side
of the Figure. From Proposition \ref{prop:cutoff_k}, it follows that
consumers with more than $\widehat{k}$ connections still search with
probability one. Suppose then that the search cost decreases gradually. This
change in search cost first affects the equilibrium search probability $q(%
\widehat{k})$ of consumers with $\widehat{k}$ connections, until this search
probability obtains an extreme value of 1. At this point, consumers with $%
\widehat{k}$ connections are indifferent between searching and not searching
only when they search with probability one, while those with more than $%
\widehat{k}$ connections still search with probability zero. The difference
in payoffs from being passive between consumers with $\widehat{k}$ friends
and those with $\widehat{k}+1$ friends is related to the number of
connections they have. This means that these two types of consumers cannot
be indifferent between searching and not searching simultaneously. Then,
obviously there is a range of search costs where consumers with $\widehat{k}$
connections strictly prefer to search, whereas those with $\widehat{k}+1$
links strictly prefer not to search (yet).

We do not explicitly perform the comparative statics exercise for this
asymmetric model of search and information acquisition, as it is similar to
the one in the previous section. In particular, it remains true that for any
positive (sufficiently) small cost $s$ price dispersion remains if the
number of connections becomes large as some consumers will be free riding.

\section{Conclusion}

\label{sec:conclusion}

In this paper we have analyzed how word-of-mouth (WOM) communication through
social networks affects information acquisition and diffusion by consumers
and how this impacts the market power of firms. Without WOM communication
our model is prone to the Diamond paradox where the market breaks down due
to the fact that no consumer makes price comparisons. WOM communication
overcomes the Diamond paradox. Consumers that do not actively search
themselves and free-ride on their friends in the social network may well be
informed about different prices. The price comparisons they make provide
positive externalities to the rest of the consumer population that actively
searches as firms compete to be able to also sell to them.

As some consumers do compare prices, while others do not, the market is
characterized by price dispersion. The level of prices and the nature of
price dispersion depends on the network architecture, the search cost and
how quickly information is diffused in the social network. In the context of
evaluating the impact of online markets and social networks it is important
to know how expected price, price dispersion and firms' profits react to (%
\textit{i}) a decrease in search cost, (\textit{ii}) an increase in the
connectivity of the social network and (\textit{iii}) an increase in the
speed of information diffusion in the network. We find that there are
opposing effects as the increased connectivity and speed of information
diffusion lower expected market prices, whereas the decrease in search cost
increases them. Importantly, price dispersion does not disappear even if all
consumers are very well connected.

We see our paper as making a first step in analyzing how WOM communication
and sequential search interact with each other. There are obvious ways our
work can be extended in different directions. One direction that may be
taken is to analyze markets with product differentiation a la \cite%
{wolinsky1986}. In this case, consumers may not only communicate about
prices, but also about the product match. In such markets, the degree of
homophily (defined as the closeness of a consumer's preferences to those of
his neighbors) will be important. Another direction for future research
would be to model the incentives to share information directly. In some
markets, consumers receive a financial benefit from firms for a successful
referral and an important question is how consumers will react to such
incentives and when such a financial incentive is optimal from a firm's
perspective and to whom to give it.

\pagebreak

\section{Appendix A: Proofs}

\begin{singlespace}

\label{sec:appendix}

\subsection*{{\protect\normalsize {Proof of Theorem \protect\ref%
{theorem:RPERG}}}}

We use the following facts to rewrite the consumer's indifference condition.
First, $E[p]=\overline{p}-\int_{\underline{p}}^{\overline{p}}F(p)dp=\eta 
\overline{p}\ln \left( 1+\frac{1}{\eta }\right) .$ Second, $\overline{p}=r=%
\frac{s}{1-\eta \ln \left( 1+\frac{1}{\eta }\right) }$. Third, $E[\min
\left\{ p_{1},p_{2}\right\} ]=\overline{p}-2\int_{\underline{p}}^{\overline{p%
}}F(p)dp+\int_{\underline{p}}^{\overline{p}}F^{2}(p)dp$ so that%
\begin{eqnarray*}
E[p]-E[\min \left\{ p_{1},p_{2}\right\} ] &=&\int_{\underline{p}}^{\overline{%
p}}F(p)dp-\int_{\underline{p}}^{\overline{p}}F^{2}(p)dp \\
&=&\eta \overline{p}\left( (1+2\eta )\ln \left( 1+\frac{1}{\eta }\right)
-2\right) .
\end{eqnarray*}%
Thus, the consumers' indifference condition can be rewritten as 
\begin{equation*}
\begin{aligned} (1-\delta)v &&=&&&(1-\delta \tau (1-q))s + \delta
\widetilde{\tau}(q)\frac{\eta \left((1+2\eta)\ln \left(
1+\frac{1}{\eta}\right) -2\right) }{1-\eta \ln
\left(1+\frac{1}{\eta}\right)}s\\ &&&&&+ (1-\delta)s \frac{\eta \ln\left(1 +
\frac{1}{\eta}\right)}{1 - \eta \ln\left(1 + \frac{1}{\eta}\right)},
\end{aligned}
\end{equation*}%
or 
\begin{equation}  \label{eq:IC(eta)}
\dfrac{s}{v}=\dfrac{1-\delta}{1-\delta \tau (1-q) + \frac{\eta}{1-\eta
\ln\left( \frac{1+\eta }{\eta }\right)} \left[\delta \widetilde{\tau }%
(q)\left( (1+2\eta )\ln \left( \frac{1+\eta }{\eta } \right)-2\right) +
(1-\delta)\ln \left(\frac{1+\eta }{\eta } \right)\right] },
\end{equation}

\noindent We know that $\eta $ is a function of $q$ and that $%
\lim_{q\uparrow 1}\eta (q)=\infty$.

As the RHS of \eqref{eq:IC(eta)} is continuous in $0<q<1$, to prove the
existence of an RPE for small enough values of $s$ it is sufficient to show
that the RHS of \eqref{eq:IC(eta)} is positive and approaches 0 zero as $%
q\uparrow 1$. As $\eta \ln \left( \frac{1+\eta }{\eta }\right) <1,$ the
denominator is clearly positive if $\ln \left( \frac{1+\eta }{\eta }\right) >%
\frac{2}{1+2\eta }$. As for $\eta \downarrow 0$ this latter inequality
clearly holds, while the LHS and the RHS both approach 0 as $\eta
\rightarrow \infty $, this inequality holds for all $\eta $ if the
derivative of the LHS is more negative than that of the RHS. The derivate of
the LHS is $-\frac{1}{\eta (1+\eta )}$, while the derivate of the RHS is $-%
\frac{2}{(1+2\eta )^{2}}$. It is easy to see that the former derivate is
smaller than the latter. Thus, as both numerator and denominator of %
\eqref{eq:IC(eta)} are positive, the whole expression is clearly positive.

Also that $1-\delta \leq 1 - \delta\tau(1-q)$ means that the numerator is
smaller than the denominator of the RHS of \eqref{eq:IC(eta)} for $0<q<1$.
Thus, there exists an upper bound $\overline{s}$ on the search cost for
which an RPE may exist.

To demonstrate the RHS of \eqref{eq:IC(eta)} converges to zero as $q\to 1$,
we employ the following evaluations: 
\begin{eqnarray}
\lim_{\eta\to \infty} \eta \ln\left(1 + \frac{1}{\eta}\right) =
\lim_{z\downarrow 0} \frac{\ln(1+z)}{z}\ \ \ 
\mathrel{\stackrel{\makebox[0pt]{\mbox{\normalfont\tiny
l'Hopital}}}{=}}\ \ \ \lim_{z \downarrow 0} \frac{1}{1+z} = 1,
\label{eq:etaln} \\
\lim_{\eta\to \infty} \frac{\eta \ln\left(1 + \frac{1}{\eta}\right)}{1 -
\eta \ln\left(1 + \frac{1}{\eta}\right)} = \frac{\lim\limits_{\eta\to
\infty} \eta \ln\left(1 + \frac{1}{\eta}\right)}{1-\lim\limits_{\eta\to
\infty} \eta \ln\left(1 + \frac{1}{\eta}\right)}= \infty,
\end{eqnarray}
and 
\begin{equation}  \label{eq:dEp}
\begin{aligned} \lim\limits_{\eta \to \infty}\frac{\eta\left( (1+2\eta )\ln \left(\frac{1+\eta }{\eta} \right)-2\right)}{1-\eta \ln\left( \frac{1+\eta }{\eta }\right)} &&=&&& \lim_{z \downarrow 0 }\frac{\frac{1}{z}\left(\left(1 + \frac{2}{z}\right)\ln(1+z)-2 \right)}{1-\frac{\ln(1+z)}{z}}\\ &&=&&& \lim_{z \downarrow 0 }\frac{(2+z)\ln(1+z)-2z}{z^2-z\ln(1+z)}\\ && \mathrel{\stackrel{\makebox[0pt]{\mbox{\normalfont\tiny
l'Hopital}}}{=}} &&& \lim_{z \downarrow 0 }\frac{\ln(1+z) + \frac{2+z}{1+z}-2}{2z - \ln(1+z) - \frac{z}{1+z}} \\ && =&&& \lim_{z \downarrow 0 }\frac{(1+z)\ln(1+z)-z}{z(1+2z)-(1+z)\ln(1+z)}\\ &&\mathrel{\stackrel{\makebox[0pt]{\mbox{\normalfont\tiny
l'Hopital}}}{=}}&&& \lim_{z \downarrow 0 }\frac{\frac{z}{(1+z)^2}}{\frac{z(3+2z)}{(1+z)^2}} = \lim_{z \downarrow 0 }\frac{1}{3+2z} = \frac{1}{3}. \end{aligned}
\end{equation}
Finally, noting $\lim\limits_{q \uparrow 1}\widetilde{\tau}(q) = 1 - 2
\tau(1/2)$, we can see that the denominator of the RHS of \eqref{eq:IC(eta)}
increases unboundedly as $q \uparrow 1$. This means that the RHS of %
\eqref{eq:IC(eta)} converges to zero as $q \uparrow 1$.

Now, we show the limiting price for $s\downarrow 0$. We know that $s$
approaching zero is associated with $q\uparrow 1$, or $\eta \rightarrow
\infty $ implying that price dispersion vanishes. Then, it suffices to
evaluate the limiting value of $r$. We note that 
\begin{equation*}
\Resize{}{ \begin{aligned} r &=&& \frac{s}{1-\eta \ln\left(1 +
\frac{1}{\eta}\right)} \\ &= &&
\frac{(1-\delta)v}{(1-\delta\tau(1-q))\left(1-\eta \ln\left(1 +
\frac{1}{\eta}\right)\right) + \eta\left[\delta \widetilde{\tau }(q)\left(
(1+2\eta)\ln \left(\frac{1+\eta }{\eta} \right)-2\right) + (1-\delta)\ln
\left(\frac{1+\eta }{\eta } \right)\right]}. \end{aligned} }
\end{equation*}%
Notice that the numerator of $r$ does not depend on $s$. Recalling %
\eqref{eq:etaln}, we can see that the first term in the denominator
converges to zero as $q\uparrow 1$ (with associated $\eta \rightarrow \infty 
$). As 
\begin{eqnarray*}
\lim\limits_{\eta \rightarrow \infty }\eta \left( (1+2\eta )\ln \left( \frac{%
1+\eta }{\eta }\right) -2\right) &=&\lim\limits_{z\downarrow 0}\frac{\left(
(1+\frac{2}{z})\ln \left( 1+z\right) -2\right) }{z} \\
&=&\lim\limits_{z\downarrow 0}\frac{\left( (z+2)\ln \left( 1+z\right)
-2z\right) }{z^{2}} \\
&\mathrel{\stackrel{\makebox[0pt]{\mbox{\normalfont\tiny
l'Hopital}}}{=}}&\lim\limits_{z\downarrow 0}\frac{\ln (1+z)+\frac{2+z}{1+z}-2%
}{2z} \\
&=&\lim\limits_{z\downarrow 0}\frac{(1+z)\ln (1+z)-z}{2z(1+z)} \\
&\mathrel{\stackrel{\makebox[0pt]{\mbox{\normalfont\tiny
l'Hopital}}}{=}}&\lim\limits_{z\downarrow 0}\frac{\ln (1+z)+1-1}{2+4z}=0,
\end{eqnarray*}%
it follows that the second term in the denominator approaches $1-\delta $ as 
$\eta \rightarrow \infty $. Then, the entire term converges to $(1-\delta
)v/(1-\delta )=v$ as $q\uparrow 1$. This completes the proof.

\bigskip

\subsection*{{\protect\normalsize {Proof of Proposition \protect\ref%
{prop:t_infty}}}}

The proof that price dispersion must remain in the limit as $t(\infty
)\rightarrow 1$ is by contradiction. Suppose to the contrary that price
dispersion vanishes in the limit, or alternatively, that $\eta \rightarrow
\infty $.

As 
\begin{equation*}
\lim_{\eta \rightarrow \infty }r=\lim_{\eta \rightarrow \infty }\frac{s}{%
1-\eta \ln (1+\frac{1}{\eta })}=\infty ,
\end{equation*}%
it follows that in an RPE $E[p]$ must also increase without bound. But this
would imply that the pay-off of becoming active is negative, which cannot be
the case in active markets. Hence, price dispersion cannot vanish as $%
k\rightarrow \infty $ (or alternatively $t(\infty )\rightarrow 1$) and $%
\lim_{t(\infty )\rightarrow 1}q=\overline{q}$, where $0<\overline{q}<1$.

We next consider the question whether an active market exists in the limit
as $t(\infty )\rightarrow 1$. To this end, rewrite the consumers'
indifference condition 
\begin{equation*}
v-E[p]-s=\delta \Big(v-E[p]+\widetilde{\tau }(q)(E[p]-E_{\min }[p])-\tau
(1-q)s\Big),
\end{equation*}%
as 
\begin{equation*}
\Resize{}{ \begin{aligned} \frac{s}{v} &&=&&& \frac{1-\delta}{1-\delta
\tau(1-q) + \left[\delta \widetilde{\tau}(q)\frac{E[p]-E_{\min}[p]}{s} +
(1-\delta)\frac{E[p]}{s}\right]}\\ &&=&&&\frac{1-\delta}{1-\delta \tau(1-q)
+ \left[\frac{\eta }{1-\eta \ln\left(1 - \frac{1}{\eta}\right)}\left\{\delta
\widetilde{\tau}(q) \left((1+2\eta)\ln\left(1 +
\frac{1}{\eta}\right)-2\right) +
(1-\delta)\ln(1+\frac{1}{\eta})\right\}\right]}, \end{aligned} }
\end{equation*}%
where the LHS is the normalized cost of being active, while the RHS
represents the incremental benefit of being so. Observe that the denominator
of the RHS consists of three terms, the third one of which is the sum in the
large square brackets. The first and the third terms are positive, while the
second term is negative. As the second term converges to zero and $%
\widetilde{\tau }(q)$ to 1 (when $t(\infty) \to 1$), the limiting
indifference condition is approximately%
\begin{equation}
\begin{aligned} \frac{s}{v} &&=&&& \frac{1-\delta}{1 + \frac{1}{s}\left[
\lim\limits_{t(\infty)\to 1}E[p] - \delta\lim\limits_{t(\infty)\to
1}E_{\min}[p]\right]}\\ &&=&&&\frac{1-\delta}{1 + \lim\limits_{t(\infty)\to
1}\frac{\eta }{1-\eta \ln\left(1 -
\frac{1}{\eta}\right)}\left[\ln(1+\frac{1}{\eta}) - 2\delta
\left(1-\eta\ln\left(1 + \frac{1}{\eta}\right)\right)\right]}. \end{aligned}
\label{eq:IC_limit}
\end{equation}

\bigskip As the RHS approaches $0$ if $\eta \rightarrow \infty $ it is clear
this equation can always be satsified if $s$ is small enough.

\subsection*{{\protect\normalsize {Proof of Proposition \protect\ref%
{prop:cs_bs}}}}

Observe that changes $s$ affect the LHS of \eqref{eq:IC(eta)} only. In
particular, the LHS is increasing in $s$. As the RHS of the indifference
equation \eqref{eq:IC(eta)} must be decreasing in $q$ in a stable RPE, the
optimal search probability $q$ must be decreasing in $s$.

\subsection*{{\protect\normalsize {Proof of Lemma }}\ref{lem:noatom_gen}}

Suppose that in equilibrium there is an atom at $\widetilde{p}$. As the
share of active consumers is strictly positive, this price $\widetilde{p}$
gets compared to another price with strictly positive probability. As there
is a strictly positive share of consumers that compare prices, undercutting $%
\widetilde{p}$ would be beneficial for firms as this yields a discontinuous
increase in the demand, a contradiction. It follows that a firm that charges
an upper bound $\overline{p}>r$ will not sell to any consumer.

\subsection*{{\protect\normalsize {Proof of Lemma 
\protect\ref{lem:uniquerho_gen}%
}}}

Observe that the LHS of (\ref{eq:rho}) is strictly decreasing in $\rho _{1}$%
. Thus, if the RHS is increasing in $\rho ,$ then if a solution to (\ref%
{eq:rho}) exists it must be a unique. From lemma \ref{lem:noatom_gen}, it 
follows that $F(p)$ must be a strictly increasing function of $p$ for $p\in 
\lbrack \underline{p},\widetilde{p}_{1}]$ where 
$\underline{p}<\widetilde{p}_{1}\leq
\min \{r,v\}$. Then, the RHS of (\ref{eq:rho}) is strictly increasing in $%
\rho $ for $\rho \in \lbrack \underline{p},\widetilde{p}_{1}]$ as 
\begin{equation*}
F(\rho )(\rho -E[p|p<\rho ])=F(\rho )\rho -\int_{\underline{p}}^{\rho
}pdF(p)=\int_{\underline{p}}^{\rho }F(p)dp
\end{equation*}%
is strictly increasing in $\rho $. (Note that we applied integration by
parts to obtain the last equality.) Next, suppose that $\widetilde{p}%
_{1}<\min \{r,v\}$ and $F(p)$ is flat in $p\in (\widetilde{p}_{1},\widetilde{%
p}_{2})$ where $\widetilde{p}_{1}<\widetilde{p}_{2}\leq \min \{r,v\}$. It
means that $F(\rho )$ and $E[p|p<\rho ]$ are constant for $\rho \in (%
\widetilde{p}_{1},\widetilde{p}_{2})$, but even then the RHS of (\ref{eq:rho}%
) is increasing in $\rho \in (\widetilde{p}_{1},\widetilde{p}_{2})$. We can
apply similar arguments to show that the RHS of (\ref{eq:rho}) is increasing
in $\rho $ for values of $\rho $ above $\widetilde{p}_{2}$ for shapes of $%
F(p)$ which have increasing and flat regions for $p\geq \widetilde{p}_{2}$.
This proves that the RHS of (\ref{eq:rho}) is indeed increasing in $\rho $,
which in turn implies that if there is $\rho $ that satisfies (\ref{eq:rho})
it must be unique.

\subsection*{{\protect\normalsize {Proof of Lemma \protect\ref{prop:r<rho}}}}

Suppose to the contrary that $\rho <r.$ First, notice that firms pricing at $%
\rho $ sell to all consumers who search them first, while if they price
slightly above $\rho $, their demand drops as these consumers choose to wait
for $p\in (\rho ,r)$ and get informed of a lower price with a strictly
positive probability, in which case they do not buy from the firm under
question. Thus, if $F(\rho )<1$, there must exist a $\underline{r}\in (\rho
,r)$ such that $F(\underline{r})=F(\rho )$. To simplify notation, let $%
E[p\leq \rho ]\equiv E[p|p\leq \rho ]$, $E_{\min }[p\leq \rho ]\equiv E[\min
\{p_{1},p_{2}\}|p_{1},p_{2}\leq \rho ]$, $E[\underline{r}\leq p\leq r]\equiv
E[p|\underline{r}\leq p\leq r]$ and $E_{\min }[\underline{r}\leq p\leq
r]\equiv E[\min \{p_{1},p_{2}\}|\underline{r}\leq p_{1}\leq r,\underline{r}%
\leq p_{2}\leq r]$. If $F(\rho )=1,$ we will say that $E[\underline{r}\leq
p\leq r]=E_{\min }[\underline{r}\leq p\leq r]=r$. It follows that the
reservation price $r$ is determined by 
\begin{equation}
\begin{aligned} s &&=&&& r - E[p]\\ &&=&&& r -
(1-F(\rho))E[\underline{r}\leq p \leq r] - F(\rho)E[p\leq \rho].
\end{aligned}  \label{eq:r_a}
\end{equation}%
As at price $r$ buyers at least weakly prefer waiting to buying, we can
write that 
\begin{equation}
(1-\delta )(v-r)<\delta \left( 1-\tau (1-\frac{q}{2})\right) s.
\label{eq:Erhopr}
\end{equation}%
In equilibrium, an individual buyer is indifferent between being active and
passive. The payoff from being active is 
\begin{equation*}
\begin{aligned} F(\rho_1)(v-E[p\leq \rho_1])+\delta
(1-F(\rho_1))(v-E[\underline{r}\leq p \leq r])\\ +\delta (1-F(\rho_1))^2
\left(1 - \tau(1-\frac{q}{2})\right)(E[\underline{r}\leq p \leq r]
-E_{\min}[\underline{r}\leq p \leq r])\\ +\delta F(\rho_1)(1-F(\rho_1))
\left(1 - \tau(1-\frac{q}{2})\right) (E[\underline{r}\leq p \leq r]-E[p \leq
\rho_1])-s. \end{aligned}
\end{equation*}%
Use \eqref{eq:r_a} to expand the first term and simplify to obtain 
\begin{equation*}
\begin{aligned} F(\rho_1)(v-E[\underline{r}\leq p \leq r]) - (r -
E[\underline{r}\leq p \leq r]) + \delta (1-F(\rho_1))(v-E[\underline{r}\leq
p \leq r])\\ +\delta (1-F(\rho_1))^2 \left(1 -
\tau(1-\frac{q}{2})\right)(E[\underline{r}\leq p \leq r]
-E_{\min}[\underline{r}\leq p \leq r])\\ +\delta F(\rho_1)(1-F(\rho_1))
\left(1 - \tau(1-\frac{q}{2})\right) (E[\underline{r}\leq p \leq r]-E[p \leq
\rho_1]). \end{aligned}
\end{equation*}%
Add and subtract $v$ and simplify to obtain 
\begin{equation*}
\begin{aligned} v-r - (1-\delta)(1-F(\rho_1))(v-E[\underline{r}\leq p \leq
r]) \\ +\delta (1-F(\rho_1))^2 \left(1 -
\tau(1-\frac{q}{2})\right)(E[\underline{r}\leq p \leq r]
-E_{\min}[\underline{r}\leq p \leq r])\\ +\delta F(\rho_1)(1-F(\rho_1))
\left(1 - \tau(1-\frac{q}{2})\right) (E[\underline{r}\leq p \leq r]-E[p \leq
\rho_1]). \end{aligned}
\end{equation*}

The payoff from being passive is 
\begin{equation*}
\Resize{}{ \begin{aligned} \delta\Bigg\{ \Bigg.v-(1-F(\rho_1))^2
\Big[E[\underline{r}\leq p\leq r] - \widetilde{\tau}(q)(E[\underline{r}\leq
p\leq r] - E_{\min}[\underline{r}\leq p\leq r])\Big]\\
-2F(\rho_1)(1-F(\rho_1))\left[\left(\tau(1-\frac{q}{2}) -
\frac{\tau(1-q)}{2}\right)E[\underline{r}\leq p\leq r] + \left(1 +
\frac{\tau(1-q)}{2}- \tau(1-\frac{1}{2})\right)E[p\leq \rho_1]\right]\\
-F^2(\rho_1)\Big[E[p\leq \rho_1] - \widetilde{\tau }(q)(E[p\leq \rho_1] -
E_{\min}[p\leq\rho_1])\Big] -\tau(1-q)s\Bigg.\Bigg\}, \end{aligned} }
\end{equation*}%
or 
\begin{equation*}
\Resize{}{ \begin{aligned} \delta\Bigg\{ \Bigg.v- \tau(1-q)s -
\left((1-F(\rho_1))^2 + 2F(\rho_1)(1-F(\rho_1))\left(\tau(1-\frac{q}{2}) -
\frac{\tau(1-q)}{2}\right) \right) E[\underline{r}\leq p\leq r] \\ -
\left(2F(\rho_1)(1-F(\rho_1))\left(1 + \frac{\tau(1-q)}{2}-
\tau(1-\frac{1}{2})\right) + F^2(\rho_1)\right)E[p\leq \rho_1]\\
+\widetilde{\tau}(q)(1-F(\rho_1))^2(E[\underline{r}\leq p\leq r] -
E_{\min}[\underline{r}\leq p\leq r]) +
\widetilde{\tau}(q)F^2(\rho_1)(E[p\leq \rho_1] -
E_{\min}[p\leq\rho_1])\Bigg.\Bigg\}, \end{aligned} }
\end{equation*}%
Add and subtract $r$, use \eqref{eq:r_a} to replace the positive $r$ and
simplify to obtain 
\begin{equation*}
\Resize{}{ \begin{aligned} \delta\Bigg\{ \Bigg.v- r + (1-\tau(1-q))s \\ +
\left(1-F(\rho_1) - (1-F(\rho_1))^2 -
2F(\rho_1)(1-F(\rho_1))\left(\tau(1-\frac{q}{2}) -
\frac{\tau(1-q)}{2}\right) \right) E[\underline{r}\leq p\leq r] \\ +
\left(F(\rho_1) - 2F(\rho_1)(1-F(\rho_1))\left(1 + \frac{\tau(1-q)}{2}-
\tau(1-\frac{1}{2})\right) - F^2(\rho_1)\right)E[p\leq \rho_1]\\
+\widetilde{\tau}(q)(1-F(\rho_1))^2(E[\underline{r}\leq p\leq r] -
E_{\min}[\underline{r}\leq p\leq r]) +
\widetilde{\tau}(q)F^2(\rho_1)(E[p\leq \rho_1] -
E_{\min}[p\leq\rho_1])\Bigg.\Bigg\}, \end{aligned} }
\end{equation*}%
However, as the multiplicative terms of both $E[\underline{r}\leq p\leq r]$
and $(-E[p\leq \rho _{1}])$ simplifies to $\widetilde{\tau }(q)F(\rho
_{1})(1-F(\rho _{1}))$, we can rewrite the expression as 
\begin{equation*}
\Resize{}{ \begin{aligned} \delta\Bigg\{ \Bigg.v- r + (1-\tau(1-q))s +
\widetilde{\tau}(q)F(\rho_1)(1-F(\rho_1))(E[\underline{r}\leq p\leq
r]-E[p\leq \rho_1]) \\
+\widetilde{\tau}(q)(1-F(\rho_1))^2(E[\underline{r}\leq p\leq r] -
E_{\min}[\underline{r}\leq p\leq r]) +
\widetilde{\tau}(q)F^2(\rho_1)(E[p\leq \rho_1] -
E_{\min}[p\leq\rho_1])\Bigg.\Bigg\}. \end{aligned} }
\end{equation*}%
Thus, the indifference condition of buyers is 
\begin{equation*}
\begin{aligned} (1-\delta)(v-r) = \delta (1-\tau(1-q))s +
(1-\delta)(1-F(\rho_1))(v-E[\underline{r}\leq p \leq r])\\ - \delta
F(\rho_1)(1-F(\rho_1))\left(\tau(1-\frac{q}{2}) -
\tau(1-q)\right)(E[\underline{r}\leq p\leq r]-E[p\leq \rho_1]) \\ - \delta
(1-F(\rho_1))^2\left(\tau(1-\frac{q}{2}) -
\tau(1-q)\right)(E[\underline{r}\leq p\leq r] - E_{\min}[\underline{r}\leq
p\leq r])\\ + \delta \widetilde{\tau}(q) F^2(\rho_1)(E[p\leq \rho_1] -
E_{\min}[p\leq\rho_1]). \end{aligned}
\end{equation*}%
As 
\begin{equation*}
\Resize{}{ \begin{aligned} (1-\tau(1-q))s =
\left(1-\tau(1-\frac{q}{2})\right)s
+\left(\tau(1-\frac{q}{2})-\tau(1-q)\right)s\\ =
\left(1-\tau(1-\frac{q}{2})\right)s +
\left(\tau(1-\frac{q}{2})-\tau(1-q)\right)\left(r-E[\underline{r}\leq p\leq
r] +F(\rho_1)(E[\underline{r}\leq p\leq r]-E[p\leq \rho_1])\right),
\end{aligned} }
\end{equation*}%
where we used \eqref{eq:r_a} to obtain the second equality, we can rewrite
the indifference condition as 
\begin{equation*}
\Resize{}{ \begin{aligned} (1-\delta)(v-r) = \delta
\left(1-\tau(1-\frac{q}{2})\right)s +
(1-\delta)(1-F(\rho_1))(v-E[\underline{r}\leq p \leq r])\\ + \delta
F(\rho_1)^2\left(\tau(1-\frac{q}{2}) - \tau(1-q)\right)(E[\underline{r}\leq
p\leq r]-E[p\leq \rho_1]) \\ + \delta
\widetilde{\tau}(q)\left(\tau(1-\frac{q}{2}) -
\tau(1-q)\right)\Big(r-E[\underline{r}\leq p\leq r] -
(1-F(\rho_1))^2(E[\underline{r}\leq p\leq r] - E_{\min}[\underline{r}\leq
p\leq r])\Big)\\ + \delta \widetilde{\tau}(q) F^2(\rho_1)(E[p\leq \rho_1] -
E_{\min}[p\leq\rho_1]). \end{aligned} }
\end{equation*}%
The LHS of the equation is equal to the LHS of \eqref{eq:Erhopr}. The RHS of
the indifference equation is certainly strictly larger than the RHS of %
\eqref{eq:Erhopr} as all the terms that are not included in \eqref{eq:Erhopr}
are nonnegative, while independent of whether $F(\rho )=1$ or $F(\rho )<1$
at least some terms are strictly positive. Thus, the indifference condition
of consumers and \eqref{eq:Erhopr} cannot hold simultaneously. This
completes the proof that it cannot be that $\rho <r$.

\bigskip 

\subsection*{{\normalsize Proof of Proposition \ref{prop:information flows}}}

If we for the time being write $\frac{q}{2\delta (1-q)}=\eta ,$ and using
the expressions for $E[p]$ and $E_{\min }[p]$ that were gien in the proof of
Theorem 1, we can rewrite the indifference condition (\ref{ind section 5}) as

\begin{equation*}
	\begin{aligned}
&(1-\delta )(v-E[p])-s =\delta (E[p]-E_{\min }[p]), \\
&(1-\delta )\left[ v-\frac{\eta s\ln \left( 1+\frac{1}{\eta }\right) }{1-\eta
\ln \left( 1+\frac{1}{\eta }\right) }\right] -s =\delta \frac{\eta s\left(
(1+2\eta )\ln \left( 1+\frac{1}{\eta }\right) -2\right) }{1-\eta \ln \left(
1+\frac{1}{\eta }\right) }, \\
&(1-\delta )v-s =(1-\delta )\frac{\eta s\ln \left( 1+\frac{1}{\eta }\right) 
}{1-\eta \ln \left( 1+\frac{1}{\eta }\right) }+\delta \frac{\eta s\left(
(1+2\eta )\ln \left( 1+\frac{1}{\eta }\right) -2\right) }{1-\eta \ln \left(
1+\frac{1}{\eta }\right) }, \\
&(1-\delta ) =\frac{s}{v}\left[ 1+(1-\delta )\frac{\eta \ln \left( 1+\frac{1%
}{\eta }\right) }{1-\eta \ln \left( 1+\frac{1}{\eta }\right) }+\delta \frac{%
\eta \left( (1+2\eta )\ln \left( 1+\frac{1}{\eta }\right) -2\right) }{1-\eta
\ln \left( 1+\frac{1}{\eta }\right) }\right] .
\end{aligned}
\end{equation*}

Further rewriting gives 
\begin{eqnarray*}
\frac{(1-\delta )\left( 1-\eta \ln \left( 1+\frac{1}{\eta }\right) \right) }{%
1-\delta \eta \ln \left( 1+\frac{1}{\eta }\right) +\delta \eta \left(
(1+2\eta )\ln \left( 1+\frac{1}{\eta }\right) -2\right) } &=&\frac{s}{v}, \\
\frac{(1-\delta )\left( 1-\eta \ln \left( 1+\frac{1}{\eta }\right) \right) }{%
1-2\delta \eta \left(1-\eta \ln \left( 1+\frac{1}{\eta }\right) \right)} 
&=&\frac{s}{v}.
\end{eqnarray*}

As the LHS is decreasing in $\eta $ and approaches $0$ as $\eta \rightarrow
\infty $, it follows that for small enough $s$, there exists a unique value
of $\eta $ that soles the indifference condition. As $\eta $ is increasing
in $q$ and $\eta \rightarrow \infty $ as $q$ approaches 1, it follows that
as $s\rightarrow 0,$ $q\rightarrow 1.$ The rest of the proof is identical to
the proof of Theorem 1 and therefore omitted.

\subsection*{{\protect\normalsize {Proof of Proposition \protect\ref%
{prop:cutoff_k}}}}

We prove the proposition with the help of two claims.

\begin{claim}
\label{claim:hatk_lower} If consumers with $1\leq\widehat{k}\leq \overline{k}
$ number of links search with strictly positive probability, all consumers
with numbers of links less than $\widehat{k}$ links (if there are such)
search with probability one.
\end{claim}

\begin{proof}
As $w$ represents the probability a consumer assigns to a neighbor actively
searching, a consumer with $\widehat{k}$ friends searches with positive
probability only if doing so is weakly better than not searching, i.e., 
\begin{equation*}  \label{eq:ICweak_after_asym}
\Resize{}{ \delta\left(v-E[p] +\left(1 + (1-w)^{\widehat{k}} - 2
\left(1-\frac{w}{2}\right)^{\widehat{k}}\right)(E[p]-E_{\min}[p]) -
(1-w)^{\widehat{k}}s\right)\leq v-E[p]-s. }
\end{equation*}
For consumers with less than $\widehat{k}$ links, searching yields the same
payoff as the RHS of the inequality, whereas not searching yields a payoff
strictly smaller than the LHS of the inequality as the LHS is increasing in $%
\widehat{k}$ for $0<w<1$. Hence, consumers with less than $\widehat{k}$
connections search for sure.
\end{proof}

\begin{claim}
\label{claim:hatk_higher} If consumers with $1\leq\widehat{k}\leq \overline{
k}$ number of links search with positive probability less than 1, all
consumers with numbers of links greater than $\widehat{k}$ links (if there
are such) do not search.
\end{claim}

\begin{proof}
The proof is analogous to the proof of Claim \ref{claim:hatk_lower}.
\end{proof}

If a consumer with $\widehat{k}$ links is indifferent between searching and
not searching, her optimal search probability lies between zero and one.
Then, from the above two claims it follows that consumers with lower than $%
\widehat{k}$ search for sure, whereas those with greater than $\widehat{k}$
links do not search at all.

\bigskip

\subsection*{{\protect\normalsize {Proof of Proposition \protect\ref%
{prop:RPE_after_asym}}}}

Some parts of the proof are similar to the proof of Theorem \ref%
{theorem:RPERG}. To avoid repetition, we omit some details here. It is clear
that to obtain the equilibrium distribution function in %
\eqref{eq:CDF_after_asym}, we should have 
\begin{equation*}
\Resize{}{ \widehat{\eta}=\dfrac{\frac{\widehat{w}}{2}+ \delta
t(\widehat{k})(1-q)\left( \left(1-\frac{w}{2}\right)^{\widehat{k}}-
\frac{(1-w)^{\widehat{k}}}{2}\right) + \delta
\sum_{k=\widehat{k}+1}^{\overline{k}}t(k)\left( \left(1-\frac{w}{2}\right)
^{k} - \frac{(1-w)^{k}}{2}\right)}{\delta t(\widehat{k})(1-q)\left(
1+(1-w)^{\widehat{k}}-2\left( 1-\frac{w}{2}\right) ^{\widehat{k}}\right)
+\delta \sum_{k=\widehat{k} +1}^{\overline{k}}t(k)\left( 1+(1-w)^{k}-2\left(
1-\frac{\widehat{w }}{2}\right) ^{k}\right) }. }
\end{equation*}

It is clear that $\widehat{\eta }$ is a function of $\widehat{k}$ and $q$
and that $\lim\limits_{q\to 1}\widehat{\eta }(\widehat{k}=\overline{k }%
,q)=\infty $.

If a consumer with $\widehat{k}$ friends is indifferent between being
passive and active, then it must be the case that the pay-off of being
passive 
\begin{equation*}
\delta\left(v-E[p] +\left(1 + (1-w)^{\widehat{k}} - 2 \left(1-\frac{w}{2}%
\right)^{\widehat{k}}\right)(E[p]-E_{\min}[p]) - (1-w)^{\widehat{k}}s\right)
\end{equation*}
is equal to the pay-off $v-E[p]-s$ of being active. We can rewrite this
indifference equation as

\begin{equation*}
\left(1 - \delta (1-w)^{\widehat{ k}}\right)s + \delta W(E[p]-E_{\min }[p])=
(1-\delta)(v-E[p]),
\end{equation*}
where $W=\left[ 1+\left( 1-w\right) ^{\widehat{k} }-2\left( 1-\frac{w}{2}%
\right) ^{\widehat{k}}\right] $ and, using the expressions for $%
E[p],E[p]-E_{\min }[p]$ and $\overline{p}$ developed in the beginning of the
proof of Theorem 1, express this condition further as

\begin{equation*}
\Resize{}{\left(1 - \delta (1-w)^{\widehat{ k}}\right)s + s\delta
W\dfrac{\widehat{\eta }\left( (1+2\widehat{\eta })\ln \left(
1+\frac{1}{\widehat{\eta }}\right) -2\right) }{1-\widehat{\eta }\ln \left(
1+\frac{1}{\widehat{\eta }}\right) } = (1-\delta)\left(v -
\dfrac{\widehat{\eta }\ln \left( 1+\frac{1}{\widehat{\eta }} \right)
}{1-\widehat{\eta }\ln \left( 1+\frac{1}{\widehat{\eta }}\right) }s \right).
}
\end{equation*}
Bringing all the terms with $s$ on one side and re-arranging gives the
condition mentioned in the Proposition.

We now show that if $s\to 0,$ $\widehat{k}=\overline{k}$ and $q( \widehat{k}%
)\uparrow 1$ hold in equilibrium. It is clear that if $s\to 0$ price
dispersion must disappear as otherwise active consumers would have an
incentive to continue searching. Suppose then that if $s\to 0,$ $\widehat{k}<%
\overline{k}.$ This would imply that for consumers with $\overline{k}$
friends the pay-off of waiting is strictly larger than the pay-off of
actively searching. However, with price dispersion disappearing if $s\to 0$,
for all $0<\delta <1$ and for every consumer (no matter how many friends she
has) this cannot be the case. If $q(\widehat{k})\ $would not converge to 1
if $s\to 0$, then $\widehat{\eta }$ would converge to a finite number and
price dispersion would persist, a contradiction.

Finally, we focus on the price level to which the price distribution
converges if $s\to 0$. As $\widehat{k}=\overline{k},$ $\widehat{\eta }$
reduces to 
\begin{equation*}
\widehat{\eta }=\frac{\frac{\widehat{w}}{2} + \delta t(\widehat{k}
)(1-q)\left( \left(1-\frac{w}{2}\right) ^{\widehat{k}}-\frac{ (1- w)^{%
\widehat{k}}}{2}\right) }{\delta t(\widehat{k})(1-q)W}
\end{equation*}
so that $\lim\limits_{q\uparrow 1}\widehat{\eta }=\infty$ as both $w$ and $%
\widehat{w}$ converge to 1. Since search behavior of consumers with $%
\overline{k}$ links is of interest in the limit, we can write 
\begin{equation}  \label{eq:r_s0_asym}
\Resize{}{ \begin{split} &\lim\limits_{s\to 0}r =\lim\limits_{s\to
0}\frac{s}{1-\widehat{\eta }\ln \left( \frac{1}{\widehat{\eta }}+1\right) }
\\ & =\lim\limits_{q\uparrow 1}\dfrac{(1-\delta)v}{\left(1 -\delta
(1-w)^{\widehat{k}}\right)\left(1-\widehat{\eta} \ln \left(1 +
\frac{1}{\widehat{\eta}}\right)\right)+ \widehat{\eta }\left[\delta W \left(
(1+2\widehat{\eta})\ln \left( \frac{1+\widehat{\eta }}{\widehat{\eta }}
\right)-2\right) + (1-\delta)\ln \left( \frac{1+\widehat{\eta
}}{\widehat{\eta }} \right)\right] }. \end{split} }
\end{equation}

To evaluate the limit, we undertake similar steps as when we evaluated the
limiting $r$ in the proof of Theorem 1. Note that the numerator of the
expression is independent of $q$. In the denominator, the first two term
converges to zero, following \eqref{eq:etaln}. The second term in the
denominator converges to $1-\delta$ since 
\begin{eqnarray*}
\lim\limits_{\widehat{\eta} \to \infty}\widehat{\eta}\left( (1+2\widehat{\eta%
} )\ln \left(\frac{1+\widehat{\eta} }{\widehat{\eta}} \right)-2\right) &=&
\lim\limits_{z \downarrow 0}\frac{\left( (1+\frac{2}{z} )\ln
\left(1+z\right)-2\right)}{z} \\
&=& \lim\limits_{z \downarrow 0}\frac{\left( (z+2)\ln
\left(1+z\right)-2z\right)}{z^2} \\
&\mathrel{\stackrel{\makebox[0pt]{\mbox{\normalfont\tiny
l'Hopital}}}{=}} & \lim\limits_{z \downarrow 0} \frac{\ln(1+z) + \frac{2+z}{%
1+z}-2}{2z} \\
&=& \lim\limits_{z \downarrow 0} \frac{(1+z)\ln(1+z) -z}{2z(1+z)} \\
&\mathrel{\stackrel{\makebox[0pt]{\mbox{\normalfont\tiny
l'Hopital}}}{=}} & \lim\limits_{z \downarrow 0} \frac{\ln(1+z) +1-1}{2+4z}=0.
\end{eqnarray*}
Summing up, the limiting $r$ converges to $\frac{(1-\delta)v}{1-\delta} = v$%
. The proof is complete.

\pagebreak 
\bibliographystyle{aer}
\bibliography{xx}{}
\end{singlespace}
\ifx\undefined\bysame
\newcommand{\bysame}{\leavevmode\hbox 
to\leftmargin{\hrulefill\,\,}}
\fi

\end{document}